\documentclass[aps,superscriptaddress,nofootinbib,12pt,tightenlines]{revtex4}

\makeatletter

\usepackage{natbib}
\usepackage{graphicx}
\usepackage{dcolumn}
\usepackage{amsmath}
\begin{document}
\bibliographystyle{agsm}
\title{An empirical behavioral model of liquidity and volatility}

\author{Szabolcs Mike}


\author{J. Doyne Farmer}


\begin{abstract}
We develop a behavioral model for liquidity and volatility based on empirical regularities in trading order flow in the London Stock Exchange.  This can be viewed as a very simple agent based model in which all components of the model are validated against real data.  Our empirical studies of order flow uncover several interesting regularities in the way trading orders are placed and cancelled.  The resulting simple model of order flow is used to simulate price formation under a continuous double auction, and the statistical properties of the resulting simulated sequence of prices are compared to those of real data.  The model is constructed using one stock (AZN) and tested on 24 other stocks.  For low volatility, small tick size stocks (called Group I) the predictions are very good, but for stocks outside Group I they are not good.  For Group I, the model predicts the correct magnitude and functional form of the distribution of the volatility and the bid-ask spread, without adjusting any parameters based on prices.  This suggests that at least for Group I stocks, the volatility and heavy tails of prices are related to market microstructure effects, and supports the hypothesis that, at least on short time scales,  the large fluctuations of absolute returns $|r|$ are well described by a power law of the form $P(|r| > R) \sim R^{-\alpha_r}$, with a value of $\alpha_r$ that varies from stock to stock.

\end{abstract}
\maketitle


\section{Motivation and background}

\subsection{Toward a more quantitative behavioral economics}

\footnote{We would like to thank the James S. McDonnell
Foundation for their Studying Complex Systems Research Award, Credit
Suisse First Boston, Barclays Bank, Bob Maxfield, and Bill
Miller for supporting this research. We would like to thank Fabrizio Lillo, Bruce Lehman, Constantino Tsallis, Adlar Kim, Laszlo Gillemot, J-P. Bouchaud and Damien Challet for useful discussions, and Marcus Daniels for technical support.  We would particularly like to thank Austin Gerig for reproducing many of these results and for providing Figure~\ref{ntot}, and to Neda Zamani for providing Figure~\ref{longMem}.}  In the last two decades the field of behavioral finance has presented many examples where equilibrium rational choice models are not able to explain real economic behavior\footnote{This may be partly because of other strong assumptions that typically accompany such models, such as complete markets.  Until we have predictive models that drop these assumptions, however, we will not know whether more realistic assumptions in rational choice models are sufficient to solve these problems.} (Hirschleifer \citeyear{Hirschleifer01}, Barberis and Thaler \citeyear{Barberis03}, Camerer \citeyear{Camerer03}, Thaler \citeyear{Thaler05}, Schleifer \citeyear{Schleifer00}).  
There are many efforts underway to build a foundation for economics directly based on psychological evidence, but this imposes a difficult hurdle for building quantitative theories.  The human brain is a complex and subtle instrument, and in a general setting the distance from psychology to prices is large.  In this study we take advantage of the fact that electronic markets provide a superb laboratory for studying patterns in human behavior.   Market participants make decisions in an extremely complex environment, but in the end these decisions are reduced to the simple actions of placing and canceling trading orders.  The data that we study contain tens of millions of records of trading orders and prices, allowing us to reconstruct the state of the market at any instant in time.  We have a complete record of decision making outcomes in the context of the phenomenon we want to study, namely price formation.  Within the domain where this model is valid, this allows us to make a simple but accurate model of the statistical properties of prices.

\subsection{Goal}

Our goal here is to capture behavioral regularities in order placement and cancellation, i.e. {\it order flow}, and to exploit these regularities to achieve a better understanding of liquidity and volatility.  The practical component of this goal is to understand statistical properties of prices, such as the distribution of price returns and the bid-ask spread.  We will use logarithmic returns $r(t) = \pi_m(t) - \pi_m(t - 1)$, where $t$ is order placement time\footnote{All results in this study are done in order placement time, i.e. we increment $t \to t + 1$ just before each order placement occurs.  There can be variable numbers of intervening cancellations.} and $\pi_m$ is the logarithmic midprice.  The logarithmic midprice $\pi_m  = 1/2(\log p_a(t) + \log p_b(t))$,  where $p_a(t)$ is the best selling price (best ask) and $p_b$ is the best buying price (best bid); on the rare occasions that we need a price rather than a logarithmic price, we will use $p = \exp (\pi_m)$.  We are only interested in the size of price movements, and not in their direction.  We will take the size of logarithmic returns $|r(t)|$ as our proxy for volatility.  Another important quantity is the bid ask spread $s(t) = \log p_a(t) -  \log p_b(t)$.  The spread is important as a benchmark for transaction costs.  A small market order to buy will execute at the best selling price, and a small order to sell will execute at the best buying price, so someone who first buys and then sells in close succession will pay the spread $s(t)$.  Our goal is to relate the magnitude and the distribution of volatility and the spread to statistical properties of order flow.  The modeling task is to understand which properties of the order flow are important for understanding prices and to create a simple model for the relationship between them. 

\subsection{Liquidity\label{liquidity}}

The model we develop here describes the endogenous dynamics of liquidity.  We define liquidity as the difference between the current midprice and the price where an order of a given size can be executed.   Previous work has shown that liquidity is typically the dominant determinant of volatility, at least for short time scales (Farmer et al. \citeyear{Farmer04b}, Weber and Rosenow \citeyear{Weber04}, Gillemot, Lillo and Farmer \citeyear{Gillemot05}).  Periods of high volatility correspond to low liquidity and vice versa.   Here we model the dynamics of the order book, i.e. we model fluctuations in liquidity, and use this to predict fluctuations in returns and spreads\footnote{Volatility in order placement time is essentially the same as in transaction time.  Transaction time volatility typically gives a close approximation to real time volatility (Gillemot, Lillo and Farmer \citeyear{Gillemot05}).}.  Thus understanding liquidity is the first and principal step to understanding volatility.


\subsection{The zero intelligence approach to the continuous double auction}

Our model is based on a statistical description of the placement and cancellation of trading orders under a continuous double auction.  This model follows in the footsteps of a long list of other models that have tried to describe order placement as a statistical process (Mendelson
\citeyear{Mendelson82}, Cohen {\it et al.} \citeyear{Cohen85}, Domowitz and Wang \citeyear{Domowitz94},  Bollerslev, Domowitz and Wang \citeyear{Bollerslev97}, Bak {\it et al.} \citeyear{Bak97},  Eliezer and Kogan \citeyear{Eliezer98}, Tang \citeyear{Tang99}, Maslov \citeyear{Maslov00}, Slanina \citeyear{Slanina01}, Challet and Stinchcombe \citeyear{Challet01}, Daniels et al. \citeyear{Daniels03}, Chiarella and Iori, \citeyear{Chiarella02}, Bouchaud, Mezard and Potters \citeyear{Bouchaud02},  Smith et al. \citeyear{Smith03}).  For a more detailed narrative of the history of this line of work, see Smith et al. (\citeyear{Smith03}). 

The model developed here was inspired by that of Daniels et al.  (\citeyear{Daniels03}).
The model of Daniels et al. was constructed to be solvable by making the assumption that limit orders, market orders, and cancellations can be described as independent Poisson processes.  Because it assumes that order placement is random except for a few constraints, it can be regarded as a zero intelligence model of agent behavior.  Although highly unrealistic in many respects, the zero intelligence model does a reasonable job of capturing the dynamic feedback and interaction between order placement on one hand and price formation on the other.  It predicts simple scaling laws for the volatility of returns and for the spread, which can be regarded as equations of state relating the properties of order flows to those of prices.  Farmer, Patelli and Zovko (\citeyear{Farmer05}) tested these predictions against real data from the London Stock Exchange and showed that, even though the model does not predict the absolute magnitude of these effects or the correct form of the distributions, it does a good job of capturing how the spread varies with changes in order flow.  The predictions for volatility are not quite as good, but are still not bad.

Despite these successes the zero intelligence model is inadequate in many respects.  Because of the unrealistic assumptions that order placement and cancellation are uniform along the price axis, to make comparisons with real data it is necessary to introduce an arbitrary interval over which order flow and cancellation densities are measured, and to assume that they vanish outside this interval.   This assumption introduces arbitrariness into the scale of the predictions and complicates the interpretation of the results.
In addition it produces price returns with non-white autocorrelations and a thin-tailed distribution that do not match real data.

\subsection{Regularities in order flow}

The model here has the same basic elements as the zero intelligence model, but each element is modified based on empirical analysis.  The model for order placement is developed in the same style as that of Challet and Stinchcombe\footnote{The order placement model of Daniels et al. assigned independent parameters for market orders and limit orders.  As we explain in more detail in Section~\ref{orderPlacement}, the model here draws all orders out of the same price distribution, generating transactions whenever the prices cross the opposite best price.  In this regard it is similar to the model of Challet and Stinchcombe (\citeyear{Challet01}).  The important difference is that all aspects of our model here are based on empirical observations.} (\citeyear{Challet01}).  In order to have a complete model for order flow we must model three things:
\begin{enumerate}
\item
The signs of orders (buy or sell) -- see Section~\ref{orderSigns}. 
\item 
The prices where orders are placed -- see Section~\ref{orderPlacement}.
\item
The frequency with which orders are cancelled -- see Section~\ref{orderCancellation}.
\end{enumerate}
In the course of modeling each of these we uncover regularities in order placement and cancellation that are interesting for their own sake.  For order placement we show that the probability of placing an order at a given price relative to the best quote can be crudely approximated by a Student distribution with less than two degrees of freedom, centered on the best quote.  We also develop a crude but simple cancellation model that depends on the position of an order relative to the best price and the imbalance between buying and selling orders in the limit order book.

The strategic motivation behind these regularities in each case are not always obvious.  Particularly for items (2) and (3), it not clear whether the regularities we observe are driven by rational equilibrium or irrational behavior.  We do not attempt to address this question here.  Instead we work in the other direction and construct a model for volatility.  Nonetheless, our studies illustrate interesting regularities in behavior that provide a intermediate milepost for obtaining any strategic understanding of market behavior.

\subsection{Method of developing and testing the model}

This model is developed on a single stock and then tested on 25 stocks.  The tests are performed by fitting the parameters of each component of the model on order flow data alone, using a simulation to make a prediction about the distribution of volatilty and spreads, and comparing the statistical properties of the simulation to the measured statistical properties of volatility and spreads in the data during the same period of time.  When we say ``prediction", we are using it in the sense of an equation of state, i.e. we are predicting contemporaneous relationships between order flow parameters on one hand and statistical properties of prices on the other. 



\subsection{Heavy tails in price returns}

Serious interest in the functional form of the distribution of prices began with Mandelbrot's (\citeyear{Mandelbrot63}) study of cotton prices, in which he showed that logarithmic price returns are far from normal and suggested that they might be drawn from a Levy distribution.   There have been many studies since then, most of which indicate that the cumulative distribution of logarithmic price changes has tails that asymptotically scale for large $| r |$ as a power law of the form $|r|^{-\alpha_r}$, where  (Fama \citeyear{Fama65},  Officer \citeyear{Officer72}, Akgiray, Booth and Loistl \citeyear{Akgiray89}, Koedijk, Schafgans and de Vries \citeyear{Koedijk90},   Loretan \citeyear{Loretan94}, Mantegna and Stanley \citeyear{Mantegna95},  Longin \citeyear{Longin96},  Lux \citeyear{Lux96},  Muller, Dacorogna and Pictet \citeyear{Muller98}, Plerou et al. \citeyear{Plerou99}, Rachev and Mittnik \citeyear{Rachev00}, Goldstein, Morris and Yen \citeyear{Goldstein04b}), but this remains a controversial topic.  The exponent $\alpha_r$, which takes on typical values in the range $2 < \alpha_r < 4$, is called the {\it tail exponent}.  It is important because it characterizes the risk of extreme price movements and corresponds to the threshold above which the moments of the distribution become infinite.  Having a good characterization of price returns has important practical consequences for risk control and option pricing.  For our purposes here we will not worry about possible asymmetries between the tails of positive and negative returns, which are in any case quite small for returns at this time scale.

From a theoretical point of view the heavy tails of price returns excite interest among physicists because they suggest nonequilibrium behavior.  A fundamental result in statistical mechanics is that, except for unusual situations such as phase transitions, equilibrium distributions are either exponential or normal distributions\footnote{For example, at equilibrium the distribution of energies is exponentially distributed and the distribution of particle velocities is normally distributed.  This is violated only at phase transitions, e.g. at the transition between a liquid and a gas.}.  The fact that price returns have tails that are heavier than this suggests that markets are not at equilibrium.  Although the notion of equilibrium as it is used in physics is very different from that in economics, the two have enough in common to make this at least an intriguing suggestion.  Many models have been proposed that attempt to explain the heavy tails of price returns (Arthur et al. \citeyear{Arthur97}, Bak, Pacuski and Shubik \citeyear{Bak97}, Brock and Hommes \citeyear{Brock99}, Lux and Marchesi \citeyear{Lux99},  Chang, Stauffer and Pandey \citeyear{Chang02}, LeBaron \citeyear{LeBaron01b}, Giardina and Bouchaud \citeyear{Giardina03}, Gabaix et al. \citeyear{Gabaix03,Gabaix06}, Challet, Marsili and Zhang \citeyear{Challet05}).  These models have a wide range in the specificity of their predictions, from those that simply demonstrate heavy tails to those that make a more quantitative prediction, for example about the tail exponent $\alpha_r$.  However, none of these models produce quantitative predictions of the magnitude and functional form of the full return distribution.  At this point it is impossible to say which, if any, of these models are correct.   


\subsection{Bid-ask spread}

In this paper we present new empirical results about the bid-ask spread.  There is a substantial empirical and theoretical literature on the spread.  A small sample is (Demsetz \citeyear{Demsetz68}, Stoll \citeyear{Stoll78}, Glosten \citeyear{Glosten88}, Glosten \citeyear{Glosten92}, Easley and O'Hara \citeyear{Easley92}, Foucault, Kadan and Kandel \citeyear{Foucault01}, Sandas \citeyear{Sandas01}).  These papers attempt to explain the strategic factors that influence the size of the spread.  We focus instead on the more immediate and empirically verifiable question of how the spread is related to order placement and cancellation. 


\subsection{Organization of the paper}

The paper is organized as follows:  Section \ref{data} discusses the market structure and the data set.  In Section~\ref{orderSigns} we review the long-memory order flow and discuss how we model the signs of orders.  In Section~\ref{orderPlacement} we study the distribution of order placement conditioned on the spread. and in Section~\ref{orderCancellation} we study order cancellation.  Section~\ref{crossSectional} we measure the parameters for the combined order flow for order signs, prices, and cancellations on all the stocks in the sample.  In Section~\ref{priceFormation} we put this together by simulating price formation for each stock based on the combined order flow model, and compare the statistical properties of our simulations to those of volatility and spreads.  Finally in the last section we summarize and discuss the implications and future directions of this work. 

\section{The market and the data\label{data}}

This study is based on data from the on-book market in the London Stock exchange.  These data contain all order placements and cancellations, making it possible to reconstruct the limit order book at any point in time.  In 1997 $57\%$ of the transactions in the LSE occurred in the on-book market and by 2002 this rose to $62\%$.  The remaining portion of the trading takes place in the off-book market, where trades are arranged bilaterally by telephone.   Off-book trades are published only after they have already taken place.  Because the on-book market is public and the off-book market is not, it is generally believed that the on-book market plays the dominant role in price formation.  We will not use any information from the off-book market here.  For a more extensive discussion of the LSE market structure, together with some comparative analysis of the two markets, see Lillo, Mike and Farmer (\citeyear{Lillo05}).

The limit order book refers to the queue that holds limit orders waiting to be executed.  The priority for executing limit orders depends both on their price and on the time when they are placed, with price taking priority over time. There are no designated market makers, though market making can occur in a self-organized way by simultaneously placing orders to buy and to sell at the same time.  The LSE on-book market is purely electronic.  Time stamps are accurate to the second.   Because we have a complete record of order placement we know unambiguously whether transactions are buyer or seller initiated.  The order book is transparent, in the sense that all orders are visible to everyone.  It  is also anonymous, in the sense that the identity of the institutions placing the orders is unknown, and remains unknown even after transactions take place.

The model that we study here was constructed based on data from the stock Astrazeneca (AZN) during the period from May 2000 - December 2002.  It was then tested on data for twenty other stocks during the same period.  Four of them had a tick size change during this period. Because this can cause important differences in behavior, we treat samples with different tick sizes separately.  As summarized in Table~\ref{dataList} there are 25 samples in all.
\begin{table}
\begin{tabular}{| c c | c c | c c |}
\hline
Stock & \# of orders & Stock & \# of orders & Stock & \# of orders\\ 
\hline
SHEL050 & 3,560,756 & BLT  & 984,251& III050  & 301,101\\
VOD & 2,676,888 & SBRY & 927,874 & TATE & 243,348\\
REED & 2,353,755 & GUS  & 836,235 & FGP & 207,390\\
AZN  & 2,329,110 & HAS  & 683,124 & NFDS & 200,654\\
LLOY & 1,954,845 & III050  & 602,416 & DEB  & 182,666\\ 
SHEL025 & 1,708,596 & BOC100  & 500,141& BSY100 & 177,286\\
PRU  & 1,413,085 & BOC050  & 345,129 & NEX & 134,991\\
TSCO & 1,180,244 & BPB & 314,414 & AVE & 109,963\\
BSY050 & 1,207,885 & & & &\\
\hline
\end{tabular}
\label{dataList}
\caption{The ticker symbols for the stocks in our data set, together with the number of orders placed during the period of the sample.  These data are all from the period from May
2, 2000 to December 31, 2002.   In cases where the tick size changes we consider the periods with different tick sizes separately.  In these cases the tick size (in hundredths of pence) is appended to the ticker symbol.} 
\end{table}

We treat the data in each sample as if it were a continuously running market.  Trading in the LSE begins with an opening auction and ends with a closing auction.  To keep things simple we remove the opening and closing auctions, and only use data during the day, when the auction is continuous.  We also remove the first hour and last half hour of each day, i.e. we consider only data from 9:00 am to 4:00 pm.  We do this because near the opening and closing auctions there are transient behaviors, such as the number of orders in the book building up and winding down, caused by the fact that many traders close out their books at the end of the day.  (This does not seem to be a large effect and does not make a great difference in our results). We paste together data from different days, ignoring everything that happens outside of the interval from 9:00 - 4:00 on trading days.  In our data analyses we are careful not to include any price movements that span the daily boundaries.  

There are several different types of possible trading orders in the LSE.   The details are not important here.  For convenience we will define an {\it effective market order} as any trading order that generates an immediate transaction, and an {\it effective limit order} as any order that does not.  A single real order may correspond to more than one effective order.  For example, a limit order that crosses the opposite best price might generate a transaction and leave a residual order in the book, which we treat as two effective orders.

\section{Generation of order signs: The important role of long-memory \label{orderSigns}}

\begin{figure}[ptb]
\begin{center}
\includegraphics[scale=0.8]{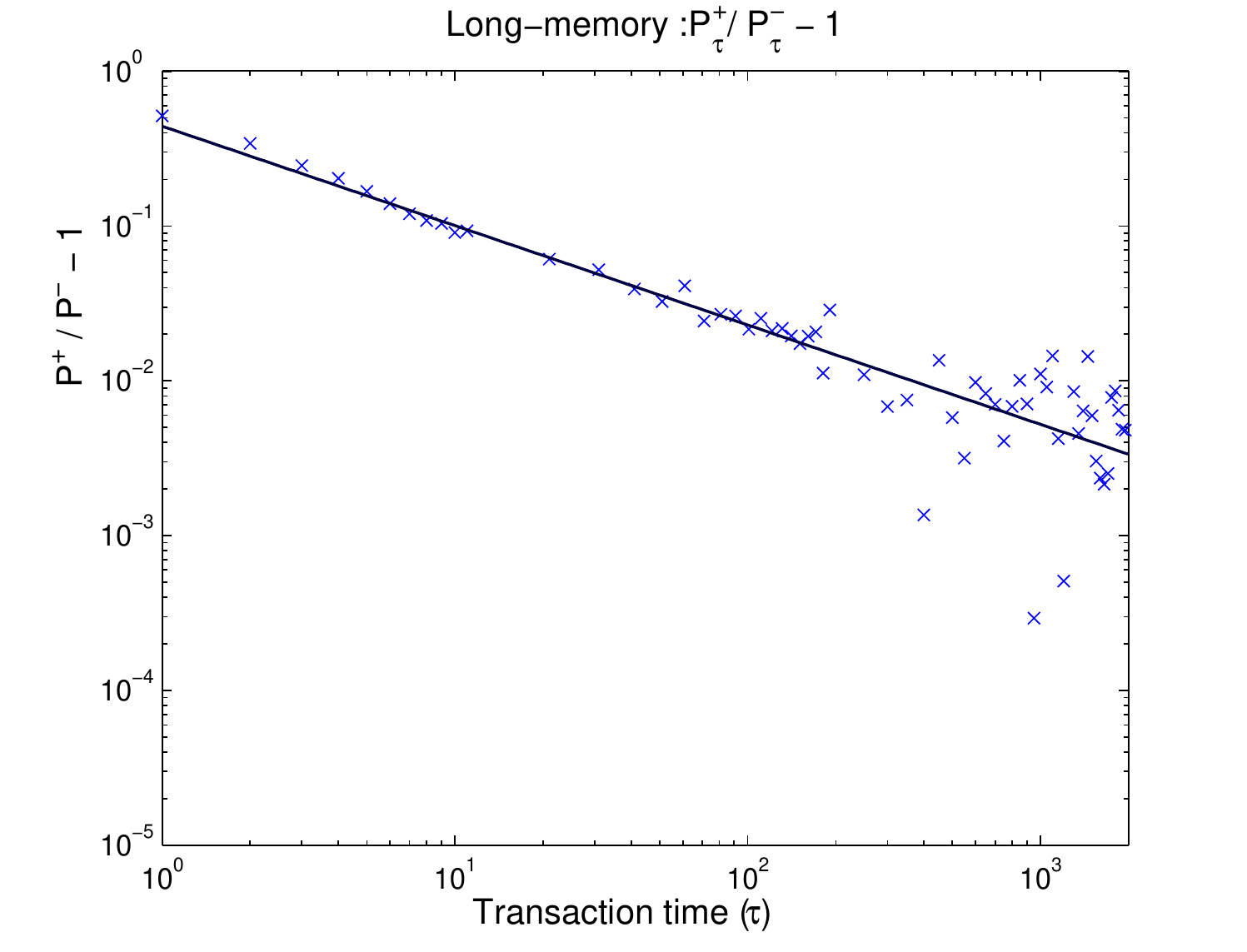}
\end{center}
\caption{An illustration of long-memory for the stock AZN.  $P_\tau^+$ is the probability that an effective market order placed at transaction time $t$ has the same sign at tune $t + \tau$, and $P_\tau^-$ is the probability that it has the opposite sign.  The crosses correspond to empirical measurements, and the line to a fitted power law $K\tau^{-\gamma}$, with $\gamma = 0.59$.}
\label{longMem}
\end{figure}

To model order placement it is necessary to decide whether each new order is to buy or to sell.  We arbitrarily designate $+1$ for buy and $-1$ for sell.  Given that returns are essentially uncorrelated in time, it might seem natural to simply assume that order signs are IID.  This is not a good approximation for the markets where this has been studied\footnote{These studies were for the Paris and London stock markets; we also observe long-memory in order signs for the NYSE, and recently Vaglica et al. have observed it in the Spanish Stock Market (\citeyear{Vaglica06}).}.  Instead, the signs of orders follow a long-memory process (Bouchaud et al. \citeyear{Bouchaud04}, Lillo and Farmer \citeyear{Lillo03c}).  Roughly speaking, this means that the autocorrelation of order signs $C(\tau)$ is positive and decays as $\tau^{-\gamma_s}$ for large $\tau$ with $0 < \gamma_s < 1$.  Because $C(\tau)$ decays so slowly, it is non-integrable.  Here $\tau$ is the time lag between the placement of two orders measured either as the intervening number of transactions; essentially the same results are obtained using elapsed clock time while the market is open\footnote{Lillo and Farmer (\citeyear{Lillo03c} showed that for the stocks they studied in the London Stock Exchange long-memory existed in both real time and transaction time, and that the differences in the values of $\gamma_s$ were statistically insignificant.}.  The coefficients of the estimated sample autocorrelation remain positive at statistically significant levels for lags of $10,000$ transactions or more, corresponding to time intervals of several weeks.  Figure~\ref{longMem}
shows an example illustrating long memory.

The observation of long-memory in order flow is surprising because it implies a high degree of predictability in order signs -- by observing the sign of an order that has just been placed, it is possible to make a statistically significant prediction about the sign of an order that will be placed two weeks later.   In order to compensate for this and keep price changes uncorrelated, the market must respond by adjusting other properties to prevent the predictability in order flow from being transmitted to the signs in price changes.  As suggested by Lillo and Farmer (\citeyear{Lillo03c}), this is achieved via a time varying liquidity imbalance, albeit with some time lag (Farmer et al., \citeyear{Farmer06}).  I.e., when effective buy market orders are likely, the liquidity for buy orders is higher than that for sell orders by a sufficient amount to make up the difference.   Alternatively, as demonstrated by Bouchaud et al. (\citeyear{Bouchaud04}, \citeyear{Bouchaud04b}) this also implies that price responses must be temporary.  We find that the long-memory properties of order signs is very important for price formation,  and strongly affects the tail exponent characterizing the distribution of large price returns.

We have proposed a model to explain the long-memory of order flow based on strategic order splitting (Lillo, Mike, and Farmer \citeyear{Lillo05b}). When an agent wishes to trade a large amount, she does not do so by placing a large trading order, but rather by splitting it into smaller pieces and executing each piece incrementally according to the available liquidity in the market.  We assume such hidden orders have an asymptotic power law distribution in their size $V$ of the form $P(V > v) \sim v^{-\beta}$, with $\beta > 0$, as observed by Gopikrishnan et al. \citeyear{Gopikrishnan00}).  Our model assumes that hidden orders enter according to an IID process, and that they are executed in constant increments at a fixed rate, independent of the size of the hidden order.  Because all the executed orders corresponding to a given hidden order have the same sign, large hidden orders cause persistence in the sequence of order signs.  We show that under these assumptions the signs of the executed orders are a long-memory process whose autocorrelation function asymptotically scales as $\tau^{- \gamma_s}$, with $\gamma_s = \beta - 1$.  This prediction is borne out empirically by comparisons of off-book and on-book  data (Lillo, Mike and Farmer, \citeyear{Lillo05b}).  

The customary way to discuss long-memory is in terms of the Hurst exponent, which is related to the exponent of the autocorrelation function as $H = 1 - \gamma/2$.  For a long-memory process the Hurst exponent is in the range $1/2 < H < 1$.   For a diffusion process with long-memory increments the variance over a period $t$ scales as $t^{2H}$, and statistical averages converge as $t^{(H - 1)}$.  This creates problems for statistical testing, as discussed in Section~\ref{priceFormation}.

For simulating price formation as we will do in Section~\ref{priceFormation} we have used the model of Lillo, Mike and Farmer described above, and we have also used a fractional gaussian random process (Beran, \citeyear{Beran94}) (in the latter case we take the signs of the resulting random numbers).   Because the algorithm for the fractional gaussian algorithm is standard and easy to implement, for purposes of reproducibility we use it for the results presented here.  As described in the next section, we first generate the sign of the order and then decide where it will be placed.  Thus we do not discriminate between effective limit orders and effective market orders in generating order signs.  This is justified by studies that we have done of the signs of effective limit orders, which exhibit long-memory essentially equivalent to that of effective market orders.

\section{Order placement \label{orderPlacement}}

\subsection{Previous studies of the order price distribution}

Even a brief glance at the data makes it clear that the probability for order placement depends on the distance from the current best prices.  This was studied in the Paris Stock Exchange by Bouchaud, Mezard and Potters (\citeyear{Bouchaud02}) and in the London Stock Exchange by Zovko and Farmer (\citeyear{Zovko02}).  Both groups studied only orders placed inside the limit order book.  For buy orders, for example, this corresponds to orders whose price is less than or equal to the highest price that is currently bid.  They found that the probability for order placement drops off asymptotically as a power law of the form $x^{-\alpha_x}$.  The value of $\alpha_x$ varies from stock to stock, but is roughly $\alpha_x \approx 0.8$ in the Paris Stock Exchange and $\alpha_x \approx 1.5$ in the London Stock Exchange.  This means that in Paris the mean of the distribution does not exist and in London the second moment does not exist.  The small values of $\alpha_x$ are surprising because they imply a significant probability for order placement even at prices that are extremely far from the current best prices, where it would seem that the probability of ever making a transaction is exceedingly low\footnote{Orders are observed at prices very far from the best price, e.g. half or double the current price.  The fact that these orders are often replaced when they expire, and that their probability of occurrence lies on a smooth curve as a function of price, suggest that such orders are intentional.}

Here we add to this earlier work by studying the probability of order placement inside the spread and the frequency of transactions conditional on the spread.  We will say that a new order is placed {\it inside the book} if its logarithmic limit price $\pi$ places it within the existing orders, i.e. so that for a buy order $\pi \le \pi_b$ or for a sell order $\pi \ge \pi_a$.   We will say it is {\it inside the spread} if its limit price is between the best price to buy and the best price to sell, i.e. $\pi_b < \pi < \pi_a$.  Similarly, if it is a buy order it generates a transaction for $\pi \ge \pi_a$ and if it is a sell order for $\pi \le \pi_b$.  To simplify nomenclature, when we are speaking of buy orders, we will refer to $\pi_b$ as the {\it same best} price and $\pi_a$ as the {\it opposite best} price, and vice versa when we are speaking of sell orders.  We will define $x$ as the logarithmic distance from the same best price, with $x = \pi - \pi_b$ for buy orders and $x = \pi_a - \pi$ for sell orders.  Thus by definition $x = 0$ for orders placed at the same best price, $x > 0$ for aggressive orders (i.e. those placed outside the book), and $x < 0$ for less aggressive orders (those placed inside the book). 

\subsection{Strategic motivations for choosing an order price}

In deciding where to place an order a trader needs to make a strategic trade off between certainty of execution on one hand and price improvement on the other.  One would naturally expect that for strategic reasons the limit prices of orders placed inside the book should have a qualitatively different distribution than those placed inside the spread.  To see why we say this, consider a buy order.  If the trader is patient she will choose $\pi < \pi_b$.   In this case the order will sit inside the limit book and will not be executed until all buy orders with price greater than $\pi$ have been removed.  The proper strategic trade off between certainty of execution and price improvement depends on the position of other orders.  Price improvement can only be achieved by being patient, and waiting for other orders to be executed.  Seeking price improvement also lowers the probability of getting any execution at all.  
In the limit where $\pi \ll \pi_b$ and there are many orders in the queue, the execution probability and price improvement should vary in a quasi-continuous manner with $\pi$, and so one would expect the probability of order placement to also be quasi-continuous.

The situation is different for an impatient trader.  Such a trader will choose $\pi > \pi_b$.  If she is very impatient and is willing to pay a high price she will choose $\pi \ge \pi_a$, which will result in an immediate transaction.  If she is of intermediate patience, she will place her order inside the spread.  In this case the obvious strategy is to place the order one price tick above $\pi_b$, as this is the best possible price with higher priority than any existing orders.  From a naive point of view it seems foolish to place an order anywhere else inside the spread\footnote{This reasoning neglects the consequences of time priority and information lags in order placement; as we will discuss later, when these effects are taken into account other values may be reasonable.}, as this gives a higher price with no improvement in priority of execution.  One would therefore naively expect to find that order placement of buy orders inside the spread is highly concentrated one tick above the current best price.  This is not what we observe.

\subsection{Our hypothesis}

To model order placement we seek an approximate functional form for $P(x | s)$, the probability density for $x$ conditioned on the spread.  This problem is complicated by the fact that for an order that generates an immediate transaction, i.e. an effective market order,  the relative price $x$ is not always meaningful.  This is because such an order can either be placed as a limit order with $x \ge s$ or as a market order, which has an effective price $x = \infty$.  Farmer et al. (\citeyear{Farmer04}) showed that for the LSE it is rare for an effective market order to penetrate deeper than the opposite best price.  The restriction to the opposite best price can be achieved either by the choice of limit price or by the choice of order size.  Thus two effective market orders with different stated limit prices may be equivalent from a functional point of view, in that they both generate transactions of the same size and price.  We resolve this ambiguity by lumping all orders with $x \ge s$ together and characterizing them by $P_\theta$, the probability that a trading order causes an immediate transaction\footnote{If only part of an order causes an immediate transaction we will treat it as two orders, one of which causes a transaction and one of which doesn't.}.  We are thus forced to try to reconstruct the probability density $P(x | s)$ using only orders with $x < s$, and then try to use this result to understand $P_\theta(s)$.

Another complication is the finite tick size $T$, the minimum increment of price change. The logarithmic price interval corresponding to one tick changes as the midprice changes.  There is a window of size one tick within which we will not see any observations inside the spread, so $P(x | s)$ is distorted within an interval $\log (p + T) - \log p \approx T/p$ of the opposite best.  Because of this, the condition for an effective limit order is more accurately written $x < s - T/p$. While these are equivalent in the limit $T \to 0$, this is not true for finite $T$, and we find that it makes a difference in our results.   

We find that we can approximate $P(x | s)$ by a density function $P^*(x)$ which is independent of the spread, as follows: 
 \begin{eqnarray}
 \label{hyp1}
 P(x | s) & = & P^*(x),  \text{  for  } -\infty < x < s - T/p\\
 \label{hyp2}
 P_\theta (s) & = & \int_{s}^{\infty} P^*(x) dx,
 \end{eqnarray}
where $P^*(x)$ is defined on $-\infty < x < \infty$.  

To understand this hypothesis it is perhaps useful to briefly explain how we will later use it to simulate order placement, as described in Section~\ref{priceFormation}.   We draw a relative price $x$ at random from $P^*(x)$.  If $x$ satisfies $-\infty < x < s - T/p$ we generate an effective limit order at logarithmic price $\pi = \pi_b + x$, and if $x \ge s - T/p$ we generate an effective market order, which creates a transaction with an order from the opposite best.   Note that with finite tick size $T$, this is equivalent to using $x > s$ as the condition for a transaction, which is why we can state Equation~\ref{hyp2} in the form that we do.  Although $P(x | s)$ is not independent of the spread, we find that the approximation above is nonetheless sufficient to generate good results in simulating the return and spread distributions.

\subsection{Method of reconstruction}

To reconstruct $P(x | s)$ for $x > 0$ we have to take account of the fact that as we vary $s$, the number of data points that satisfy the condition $x < s - T/p$ varies, so the proper normalization of the conditional distribution also varies.  The number of data points satisfying this condition is $N(s - T/p > x_j) \equiv \sum_i I(s_i - T/p_i > x_j)$, where $I(y)$ is the indicator function, which satisfies $I(y) = 0$ when $y < 0$ and $I(y) = 1$ when $y \ge 0$.  Under the assumption that $P(x | s)$ is independent of $s$ for $x < s - T/p$, we can combine data for different values of the spread by assigning each point $x_j$ a weight $w_j = N/N(s - T/p > x_j)$, where $N$ is the total number of data points in the full sample.  We can then estimate $P^*(x)$ by assigning bins along the $x$ axis and computing the average weight of the points inside each bin.  We can test for dependence of $P(x | s)$ on the spread in the region $0 < x < s_0$ by performing this analysis for a subsample of the data satisfying the condition $s > s_0$.

We also perform some data filterings that are intended to exclude cases where there are possible data errors or where people may be acting on stale information.  To avoid data errors we reject situations where the order size is greater than one million shares, and where the spread is negative or is greater than $100$ ticks.  There are only a few cases that satisfy these conditions.  More important, this data set has problems because orders placed within a given second are not guaranteed to be correctly time sequenced within that second; to avoid this we only allow orders that are the only ones placed in a given second.  To avoid cases when traders might be operating on stale information, we rejected limit orders that were placed less than $5$ seconds after any increase of the spread.  This is to prevent situations in which a large spread opens, moving the the opposite best price away, and then an order is placed at the previous best price.  With up-to-date information this order would have generated a transaction, but because of a slow response it becomes an effective limit order and remains in the book.

\subsection{Empirical test of the hypothesis}

\begin{figure}[ptb]
\begin{center}
\includegraphics[scale=0.5]{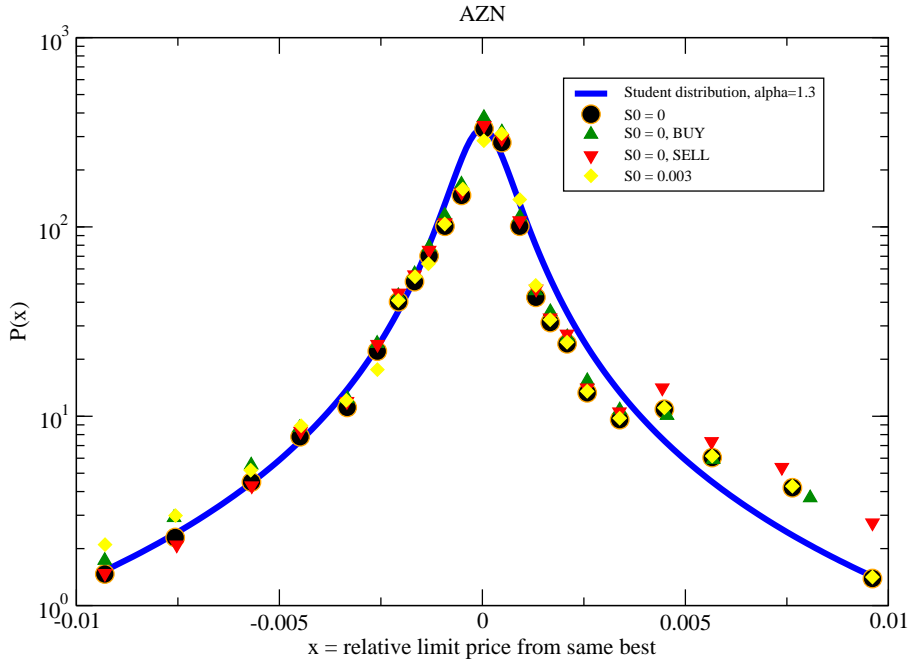}
\end{center}
\caption{Reconstruction of the probability density function $P^*(x)$ describing limit order prices as a function of $x$, the limit price relative to the same best price.  The reconstruction is done both for buy orders (green upward pointing triangles) and sell orders (red downward pointing triangles), and for two different spread conditions.  $s > s_0 = 0$ allows all $410,000$ points that survive the data filterings described in the text and that satisfy the condition $x \le s - T/p$; there are $211,000$ buy orders and $199,000$ sell orders.  There are only $26,000$ points that satisfy $s > s_0 = 0.003$.  The fitted blue curve is a Student distribution with 1.3 degrees of freedom.}
\label{collapsePricePlot}
\end{figure}

In Figure~\ref{collapsePricePlot} we show the results of reconstructing $P^*(x)$.  We use two different spread conditions, $s > s_0 = 0$ (which includes all the data) and $s > s_0 = 0.003$, and we also separate the data for buy and sell orders.  To fit $P^*(x)$ we use a generalized Student distribution\footnote{This form was suggested to us by Constantino Tsallis.  It is a functional form that is ubiquitous in the theory of non-extensive statistical mechanics (Tsallis \citeyear{Tsallis88}, Gell-Mann and Tsallis \citeyear{Gell-Mann04}).}.  The method of fitting parameters is described in White (\citeyear{White06}).  The fit is quite good for $x < 0$ and not as good for $x > 0$; in particular, it is clear that the distribution is right skewed, i.e. it has heavier tails for $x > 0$.  The data for $x > 0$ also have more fluctuations due the fact that the spread probability $P(s)$ decreases for large $s$ (see Figure~\ref{returnDist}), so there are less and less data that satisfy the condition $x < s - T/p$.  For example, for $s > s_0 = 0$ the second to left-most bin has $2600$ points, while the second to right-most bin has only $28$ points.
  
Varying $s_0$ allows us to test for independence of the spread, at least over a restricted range.  Comparing $s > s_0 = 0$ and $s > s_0 = 0.003$, the results are guaranteed to be the same for $x > s_0$, but this is not true for $0 < x < s_0$, where they will be the same only if $P(x | s)$ is independent of the spread for $0 < s < s_0$.  There are some differences, and these differences are almost certainly statistically significant, but this plot suggests that this is nonetheless not a bad approximation.   The results when buy and sell orders are separated are roughly the same.

\subsection{Predicting the probability of a transaction}

\begin{figure}[ptb]
\begin{center}
\includegraphics[scale=0.5]{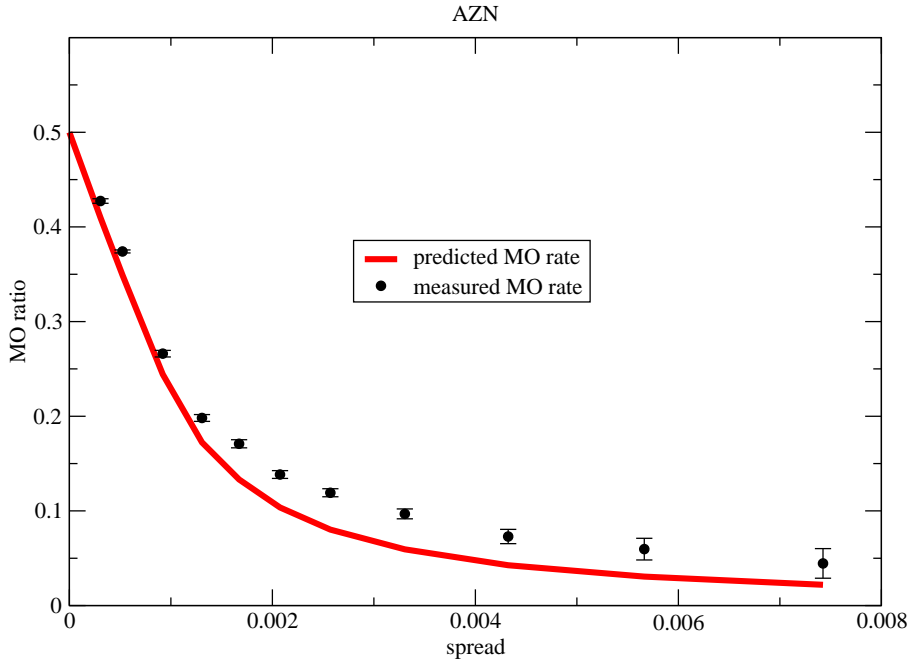}
\end{center}
\caption{The transaction probability $P_\theta$ as a function of the spread.  The curve is based on the fit to a student distribution for $P^*(x)$ in Figure~\ref{collapsePricePlot} and Equation~\ref{hyp2}.  The fraction of orders that result in transactions approaches $1/2$ in the limit as the spread goes to zero and approaches zero in the limit as the spread becomes large.}
\label{transactionRatio}
\end{figure}

We can test Equation~\ref{hyp2} using the fit to the Student distribution from Figure~\ref{collapsePricePlot}.  In Figure~\ref{transactionRatio} we plot the fraction of orders that result in transactions as a function of the spread based on Equation~\ref{hyp2}, and averaging over the midpoint prices $p$ associated with each spread.  This gives a crude fit to the data -- although the predicted transaction probabilities are generally too low, they agree well for small spreads and never differ by more than a factor of two.  The probability that an order generates a transaction approaches one half in the limit as the spread goes to zero, and approaches zero in the limit as the spread becomes large.

\section{Order cancellation \label{orderCancellation}}

In this section we develop a model for cancellation.  Cancellation of trading orders plays an important role in price formation.   It causes changes in the midprice when the last order at the best price is removed, and can also have important indirect effects when it occurs inside the limit order book.  It affects the distribution of orders in the limit order book, which can later affect price responses to new market orders.  Thus it plays an important role in determining liquidity.

\begin{figure}[ptb]
\begin{center}
\includegraphics[scale=0.5]{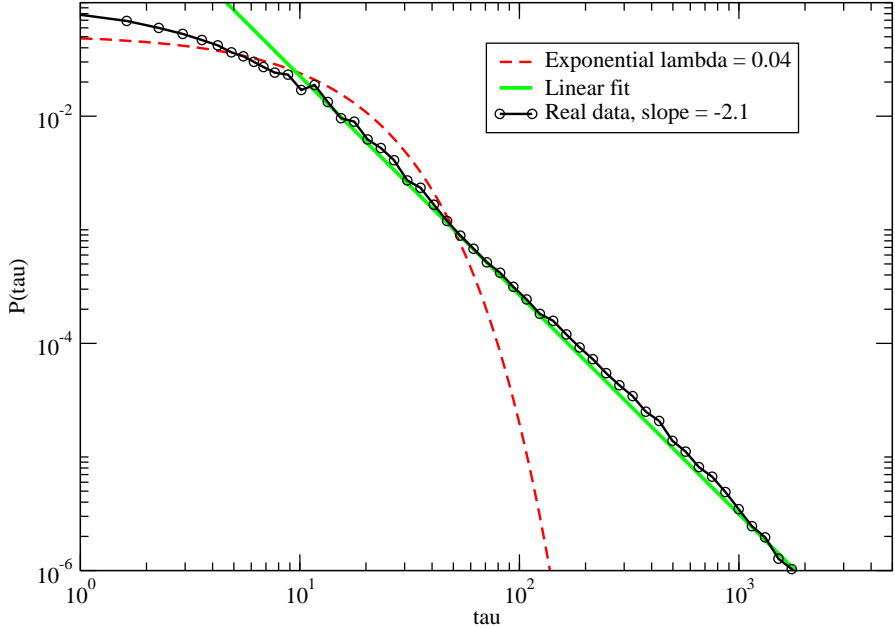}
\caption{The empirical probability density of the lifetime $\tau$ of cancelled orders for the stock Astrazeneca (black).  $\tau$ is the number of trading orders placed between the time given order is placed and the time it is cancelled. This is compared to an exponential distribution with $\lambda = 0.03$ (red).  A power law $\tau^{-(1 + \gamma_c)}$ with $\gamma_c = 1.1$ is shown for comparison.  Note that to avoid end of day effects we exclude orders that are are not cancelled between 9:00 am and 4:00 pm on trading days (but we do include orders that are placed on one day and cancelled on another day).}
\label{lifetimes}
\end{center}
\end{figure}

The zero intelligence model of Daniels et al. (\citeyear{Daniels03}) used the crude assumption that cancellation is a Poisson process.   Let $\tau$ be the lifetime of an order measured from when it is placed to when it is cancelled, where (as elsewhere in this paper), time is measured in terms of the number of intervening trading orders\footnote{Recall that we exclude orders placed at the auctions and at the beginning and end of the day.  We do not count these orders in measuring $\tau$.}.
Under the Poisson assumption the distribution of lifetimes is an exponential distribution of the form $P(\tau) = \lambda (1 - \lambda)^{\tau - 1}$.  The cancellation rate $\lambda$ can be written $\lambda = 1/E[\tau]$, where $E[\tau]$ is the expected lifetime of an order.  For AZN, for example, $\lambda \approx 0.04$.  A comparison of the exponential to the true distribution as shown in  Figure~\ref{lifetimes} makes it clear that the
Poisson process is a poor approximation of the true behavior.  The tail of the empirical density function behaves roughly like a power law of the form $\tau^{-(\gamma_c + 1)}$.   For Astrazenca $\gamma_c \approx 1.1$, and the power law is a good approximation over roughly two orders of magnitude\footnote{Power law tails in the cancellation process with a similar exponent were previously observed in Island data by Challet and Stinchcombe (\citeyear{Challet02}).}. Similar results are observed for the other stocks we studied with $1 < \gamma_c < 1.5$.  The heavy tailed behavior implies that the most long-lived orders observed in a sample of this length last an order of magnitude longer than they would under the Poisson hypothesis.   The cancellation rate  $\lambda(\tau)$ is a decreasing function of time and also depends on the identity of the order $i$.  Both of these effects contribute to generating heavy tails in the lifetime distribution of the whole population.

To reproduce the correct distribution of lifetimes, the challenge is to find a set of factors that will automatically induce the right overall time dependence $\lambda(\tau)$.  We find three such factors:  position in the order book relative to the best price, imbalance of buy and sell orders in the book, and the total number of orders.  We now explore each of these effects in turn.

\subsection{Position in the order book}

Strategic considerations dictate that position in the order book should be important in determining the cancellation rate.   Someone who places an order inside the spread likely has a very different expected execution time than someone who places an order inside the book.  If an order is placed at the best price or better, this implies that the trader is impatient and likely to cancel the order quickly if it is not executed soon.   In contrast, no one would place an order deep inside the book unless they are prepared to wait a long time for execution.  Dependence on cancellation times with these basic characteristics was observed in the Paris Stock Market by Potters and Bouchaud (\citeyear{Potters03}).

To study this effect we measure the cancellation rate as a function of the distance to the opposite best price.  Letting $\pi$ be the logarithmic price where an order is placed,  the distance of the price of the order from the opposite best at time $t $ is $\Delta_i(t) = \pi - \pi_b(t)$ for sell orders and $\Delta_i(t) = \pi_a(t) - \pi$ for buy orders.  Thus by definition $\Delta(0)$ is the distance to the opposite best when the order is placed, and $\Delta(t) = 0$ if and when the order is executed.  We compute the sample correlation $\rho(\Delta(0),\tau)$, and find that $0.1 < \rho < 0.35$ for the stocks we studied, confirming the positive association between distance to the opposite best and cancellation time.

Strategic considerations suggest that cancellation should depend on $\Delta (t)$ as well as $\Delta(0)$.   If $\Delta(t) \gg \Delta(0)$ then this means that the opposite best price is now much further away than when the order was originally placed, making execution unlikely and making it more likely that the order will be cancelled.  Similarly, if $\Delta(t) \ll \Delta(0)$ the opposite best price is quite close, execution is very likely and hence cancellation should be less likely.  This is confirmed by fact that for buy cancellations we observe positive correlations with the opposite best price movements in the range of $20 - 25\%$, and for sell orders we observe negative correlations of the same size.  In the interest of keeping the model as simple as possible we define a variable that encompasses both the dependence on $\Delta(0)$ and the dependence on $\Delta(t)$, defined as their ratio
\[
y_i(t) = \frac{\Delta_i(t)}{\Delta_i(0)}.
\]
By definition when order $i$ is placed $y_i = 1$, and if and when it is executed, $y_i = 0$.  A change in $y_i(t)$ indicates a movement in the opposite price, measured in units whose scale is set by how far from the best price the order was originally placed.  

\begin{figure}[ptb]
\begin{center}
\includegraphics[scale=0.4]{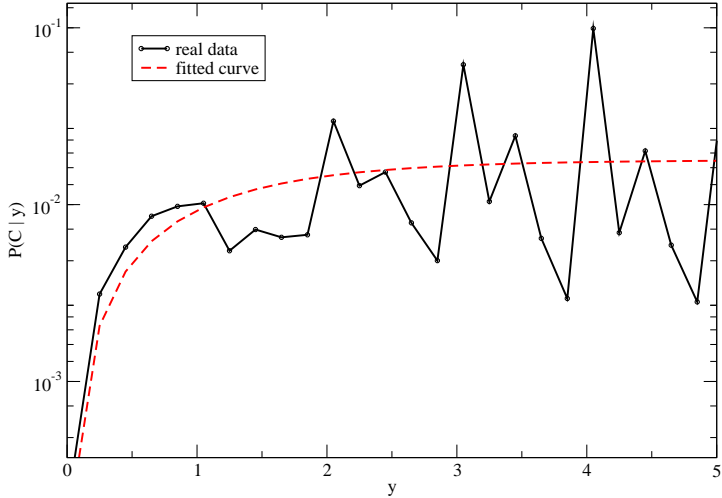}
\caption{The probability of cancellation $P(C_i | y_i)$ for AZN conditioned on $y_i(t) = \Delta_i(t)/\Delta_i(0)$.  The variable $y_i$ measures the distance from order $i$ to the opposite best price relative to its value when the order was originally placed.  The solid curve is the empirical fit $K_1(1-e^{-y_i})$, with $K_1 \approx 0.012$.}
\label{yProb}
\end{center}
\end{figure}

To measure the conditional probability of cancellation we use Bayes' rule.  The probability of canceling an individual order conditioned on $y_i$ can be written 
\begin{equation}
P(C_i | y_i) = \frac{P(y_i | C_i)}{P(y_i)} P(C),
\label{bayes}
\end{equation}
where $C_i$ is a variable that is true when the given order is cancelled and false otherwise.  $P(C)$ is the unconditional probability of canceling an order. The conditional probability $P(y_i | C_i)$ can be computed by simply making a histogram of the values of $y_i$ when cancellations occur.  Figure~\ref{yProb} shows an empirical estimate of the conditional probability of cancellation for AZN computed in this way.  
Although there are substantial oscillations\footnote{We believe these oscillations are caused by round number effects in order placement and cancellation.}, as predicted by strategic considerations, the cancellation probability tends to increase with $y_i$.  As $y_i$ goes to zero the cancellation probability also goes to zero, and it increases to a constant value of roughly $3\%$ per unit time as $y_i$ gets large (we are measuring time in units of the number of trading orders that are placed).  To approximate this behavior for modeling purposes we empirically fit a function of the form $K_1(1 - \exp(-y_i))$.  For AZN minimizing least squares gives $K_1 \approx 0.012$. 

The question remains whether the ratio $\Delta_i (t)/\Delta _i (0)$ fully captures the cancellation rate, or whether the numerator and denominator have separate effects that are not well modeled by the ratio.  To test this we divided the data into four different bins according to $\Delta_i (0)$ and repeated the measurement of Figure~\ref{yProb} for each of them separately.  We do not get a perfect collapse of the data onto a single curve.  Nonetheless, each of the four curves has a similar shape, and they are close enough that in the interest of keeping the model simple we have decided not to model these effects separately. 

\subsection{Order book imbalance\label{imbalance}} 

\begin{figure}[ptb]
\begin{center}
\includegraphics[scale=0.5]{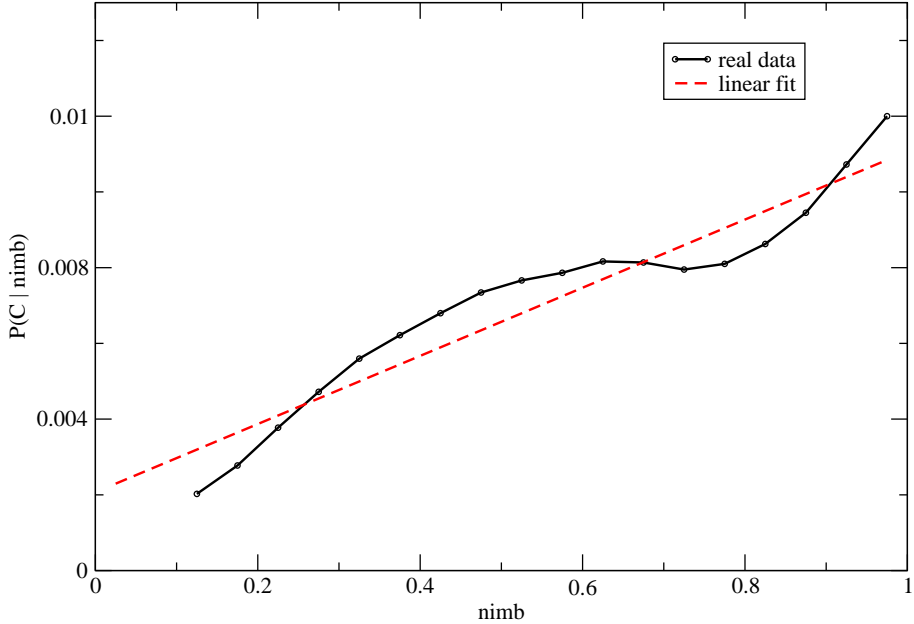}
\caption{The probability of canceling a given order, $P(C_i | n_{imb})$ for the stock AZN.  This is conditioned on the order imbalance $n_{imb}$.  The dashed curve is a least squares fit to a linear function, $K_2(n_{imb} + B)$, with $K_2 \approx 0.0098$ and $B \approx 0.20$.}
\label{nimb}
\end{center}
\end{figure}

The imbalance in the order book is another factor that has a significant effect on order cancellation.   We define an indicator of order imbalance for buy orders as $n_{imb} = n_{buy}/(n_{buy} + n_{sell}) = $ and for sell orders as $n_{imb} = n_{sell}/(n_{buy} + n_{sell})$, where $n_{buy}$ is the number of buy orders in the limit order book and $n_{sell}$ is the number of sell orders.  In Figure~\ref{nimb} we show an empirical estimate of the conditional distribution $P(C_i | n_{imb})$, defined as the probability of canceling a given order.
$P(C_i | n_{imb})$ is less than $1\%$ when $n_{imb} = 0.1$ and about $4\%$ when $n_{imb} = 0.95$, increasing by more than a factor of four.  This says that it is more likely for an order to be cancelled when it is the dominant order type on the book.  For example if the book has many more buy orders than sell orders, the probability that a given buy order will be cancelled increases (and the probability for a given sell order to be cancelled decreases).  Since the functional form appears to be a bit complicated, as a crude approximation we fit a linear function of the form $P(C_i | n_{imb}) = K_2( n_{imb} + B)$.  Minimizing least squares gives $K_2 \approx 0.0098$ and $b \approx 0.20$ for AZN.

\subsection{Number of orders in the order book\label{cancellationNumberOrders}}

\begin{figure}[ptb]
\begin{center}
\includegraphics[scale=0.5]{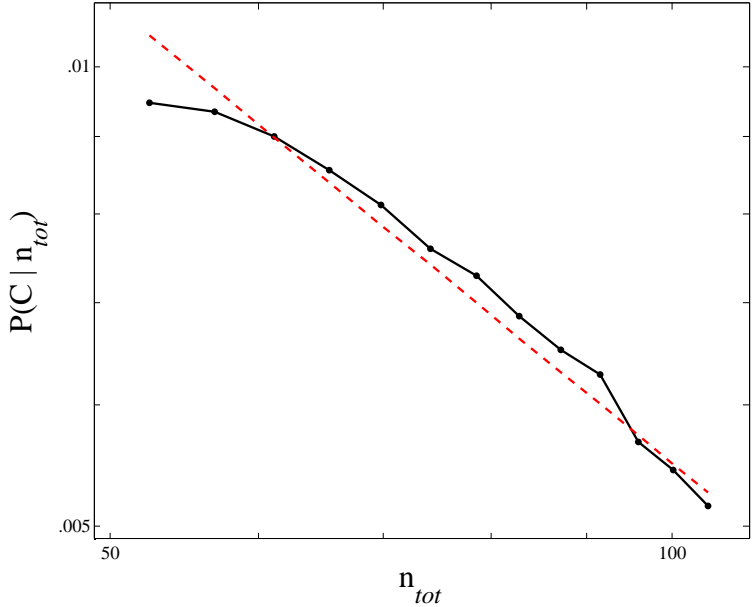}
\end{center}
\caption{The probability of canceling a given order, $P(C_i | n_{tot})$, for the stock AZN, conditioned on the total number of orders in the order book, $n_{tot}$ on a log-log plot.  The dashed line is the function $K_3/n_{tot}$, shown for reference, where $K_3=0.54$.\label{ntot}}
\end{figure}

Another variable that we find has an important effect on cancellation is $n_{tot}$, the total number of orders in the order book.  Using a procedure similar to those for the other two variables, in Figure~\ref{ntot} we plot the cancellation probability as a function of $n_{tot}$.
Surprisingly, we see that the probability of cancellation decreases as $n_{tot}$ increases, approximately proportional to $1/n_{tot}$.  A least squares fit of $\log P(C_i | n_{tot})$ vs. $b - a \log n_{tot}$ gives a slope $a = 0.92 \pm 0.06$ (using one standard deviation error bars).  The coefficient $a$ is sufficiently close to one that we simply make the approximation in our model that $P(C_i | n_{tot}) \sim 1/n_{tot}$.  We plot a line of slope $-1$ in the figure to make the validity of this approximation clear.

This is very surprising, as it indicates that the total cancellation rate is essentially independent of the number of orders in the order book.  This raises the question of how the total number of orders in the order book can remain bounded.  See the discussion in Section~\ref{stability}. 



\subsection{Combined cancellation model} 

We assume that the effects of $n_{imb}$, $y_i$, and $n_{tot}$ are independent, i.e. the conditional probability of cancellation per order is of the form
\begin{equation}
P(C_i | y_i, n_{imb}, n_{tot}) = \frac{P(y_i | C_i) P(n_{imb} | C_i) P(n_{tot} | C_i)}{P(y_i)P(n_{imb})P(n_{tot})} P(C) = A(1-\exp^{-y_i})(n_{imb}+ B)/n_{tot},
\label{cancModel}
\end{equation}
where for AZN $A = (K_1 K_2 K_3)/P(C)^2$.  For AZN $P(C) \approx 0.0075$, which together with the previously measured values of $K_1$, $K_2$, and $K_3$ gives $A \approx 1.12$. From Section~\ref{imbalance} $B \approx 0.20$.

\begin{figure}[ptb]
\begin{center}
\includegraphics[scale=0.5]{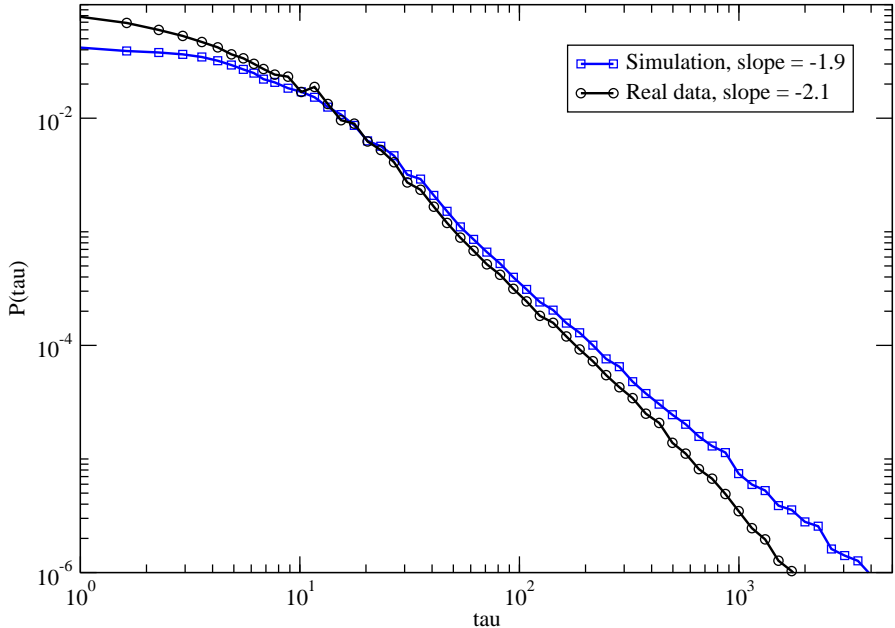}
\caption{A comparison of the distribution of lifetimes of simulated cancellations (blue squares) to those of true cancellations (black circles).}
\label{lifetimeComp}
\end{center}
\end{figure}

To test the combined model we simulate cancellations and compare to the real data.  Using the real data, after the placement of each new order we measure $y_i$, $n_{imb}$, and $n_{tot}$ and simulate cancellation according to the probability given by Equation~\ref{cancModel}.   We compare the distribution of lifetimes from the simulation to those of the true distribution in Figure~\ref{lifetimeComp}. 
The simulated lifetime distribution is not perfect, but it is much closer to the true distribution than the Poisson model (compare to Figure~\ref{lifetimes}).  It reproduces the power law tail, though with $\gamma_c \approx 0.9$, in comparison to the true distribution, which has $\gamma_c \approx 1.1$.  For small values of $\tau$ the model underestimates the lifetime probability and for large values of $\tau$ it overestimates the probability.  As an additional test of the model we plotted the average number of simulated cancellations against the actual number of cancellations for blocks of 50 events, where an event is a limit order, market order, or cancellation.   As we would hope the result is close to the identity.  Since the resulting plot is uninteresting we do not show it here.

\section{Cross-sectional study of order flow\label{crossSectional}}

The models developed in the previous three sections for order sign, price, and cancellation were constructed using only data for the stock AZN.  We then assumed that the same functional forms are valid for all the other $24$ samples described in Section~\ref{data}, and fit the parameters for the model for order sign generation developed in Section~\ref{orderSigns}, the model for order prices developed in Section~\ref{orderPlacement}, and the model for order cancellation developed in Section~\ref{orderCancellation}.  Each of the three sub-models are fit completely independently; their five independent parameters plus the tick size fully specifies the model of order flow.   The results for each of our 25 samples are summarized in Table~\ref{paramSummary}.
\begin{table}[htdp]
\begin{center}
\begin{tabular}{c|c|c|c|c|c|c}
stock ticker & $H_s$ & $\alpha_x$ & $\sigma_x \times10^{-3}$ & A & B & T \\
\hline
AVE & 0.88 & 1 & $2.7$ & 1.11 & 0.22 & 0.25\\
\hline
AZN & 0.77 & 1.31 & $2.4$ & 1.12 & 0.20 & 1 \\
\hline
BLT & 0.85 & 1.15 & $2.5$ & 1.19 & 0.19 & 0.25\\
\hline
BOC050 & 0.85 & 1.45 & $2.1$ & 1.21 & 0.2 & 0.5\\
\hline
BOC100 & 0.84 & 1.65 & $2.6$ & 1.16 & 0.18 & 1\\
\hline
BPB & 0.88 & 1.22 & $2.6$ & 1.09 & 0.21 & 0.25\\
\hline
BSY050 & 0.8 & 1.12 & $2.5$ & 1.11 & 0.19 & 0.5\\
\hline
BSY100 & 0.75 & 1.23 & $2.3$ & 0.95 & 0.18 & 1\\
\hline
DEB & 0.85 & 1.12 & $2.5$ & 1.23 & 0.20 & 0.25\\
\hline
FGP & 0.84 & 1.26 & $2.5$ & 1.03 & 0.20 & 0.25\\
\hline
GUS & 0.80 & 1.10 & $2.5$ & 1.10 & 0.20 & 0.5\\
\hline
HAS & 0.82 & 1.10 & $2.4$ & 0.99 & 0.21 & 0.25\\
\hline
III050 & 0.8 & 1.21 & $2.8$ & 1.04 & 0.23 & 0.5\\
\hline
III100 & 0.80 & 1.15 & $2.0$ & 1.24 & 0.21 & 1\\
\hline
LLOY & 0.81 & 1.25 & $2.6$ & 0.89 & 0.22 & 0.5\\
\hline
NEX & 0.85 & 1.31 & $2.1$ & 1.12 & 0.21 & 0.5\\
\hline
NFDS & 0.85 & 1.21 & $2.5$ & 1.05 & 0.20 & 0.25\\
\hline
PRU & 0.8 & 1.12 & $2.4$ & 1.02 & 0.2 & 0.5\\
\hline
REED & 0.87 & 1.09 & $2.4$ & 0.98 & 0.23 & 0.5\\
\hline
SBRY & 0.80 & 1.14 & $2.6$ & 1.01 & 0.21 & 0.25\\
\hline
SHEL025 & 0.88 & 1.43 & $2.2$ & 1.54 & 0.2 & 0.25\\
\hline
SHEL050 & 0.88 & 1.50 & $2.4$ & 1.50 & 0.2 & 0.5\\
\hline
TATE & 0.85 & 1.23 & $2.6$ & 1.12 & 0.18 & 0.25\\
\hline
TSCO & 0.82 & 1.22 & $2.2$ & 0.83 & 0.18 & 0.25\\
\hline
VOD & 0.80 & 1.05 & $2.8$ & 0.73 & 0.19 & 0.25\\
\end{tabular}
\caption{The measured parameters of our order flow models.  The first column contains the ticker symbol for the stock; where there are tick size changes we have appended the tick size.  $H_s$ is the Hurst exponent of the order sign series, $\sigma_x$ and $\alpha_x$ are the scale parameter and degrees of freedom of the order placement distribution, and T is the tick size.  The probability of cancellation for a given order is $P(C_i | y_i, n_{imb}, n_{tot}) = A(1-e^{-y_i})(n_{imb}+B)/n_{tot}$.\label{paramSummary}}
\end{center}
\end{table}

The first column of Table~\ref{paramSummary} is the Hurst exponent $H$ of the sequence of signs of limit and market orders, which characterizes the long-memory of supply and demand as discussed in Section~\ref{orderSigns}. The estimates are based on the DFA method with polynomials of degree one (Peng et al. \citeyear{Peng94}).  This summarizes the degree of long-memory in the flow of supply and demand.  The measured values are in the range $0.75 \leq H_s \leq 0.88$, a variation of roughly $15\%$.  The results are consistent with those of Lillo and Farmer (\citeyear{Lillo03c}).

The second and third columns are the tail exponent $\alpha_x$ and the scale parameter $\sigma_x$ for the Student distribution that characterizes the probability of choosing the price of an order relative to the best price for orders of the same sign, as described in Section~\ref{orderPlacement}.  The tail exponents are in the range $1 \leq \alpha_x \leq 1.65$, a variation of about $50\%$, and the scale parameters are in the range $2.0 \times 10^{-3} \leq \sigma_x \leq 2.8 \times 10^{-3}$, a variation of about $30\%$.

The fourth and fifth columns are the two parameters $A$ and $B$ that characterize the rate of order cancellation, as described in Section~\ref{orderCancellation}.  $A$ is in the range $0.73 \leq A \leq 1.54$, a variation of about $70\%$, and B is in the range $0.18 \leq B \leq 0.23$, a variation of about $25\%$.  

Finally the last column is the tick size for the sample measured in pence, which is determined by the exchange and remains constant throughout each sample.  The possible tick sizes are $0.25$, $0.5$, and $1$ pence.  

We have not attempted to compute error bars in Table~\ref{paramSummary} for two reasons.  First, because of the long-memory of both the order signs and the relative position $x$ for order placement, they are difficult to compute; see the discussion in Section~\ref{simulatedVsReal}.  Second, while the variation of parameters from stock to stock might be interesting for its own sake, our main purpose here is to perform the simulations of liquidity dynamics and volatility described in the next section, and we perform a statistical analysis there.  It is clear from this study that at least some of the parameters exhibit statistically significant variations from sample to sample.

\section{Liquidity and volatility\label{priceFormation}}

The order flow model summarized above can be used to simulate the dynamics of the limit order book.  The result is a model for the endogenous liquidity dynamics of the market.  Order placement and cancellation are modeled as conditional probability distributions, with conditions that depend on observable variables such as the number of orders in the order book.  As orders arrive they affect the best prices, which in turn affects order placement and cancellation.  This makes it possible to simulate a price sequence and compare its statistical properties to those of the real data.
 
\subsection{Description of the price formation model}

To simulate price formation we make some additional simplifying assumptions.
\begin{itemize}
\item
{\it All orders have constant size}.  This is justified by our earlier study of the on-book market of the London Stock Exchange in (Farmer \citeyear{Farmer04} et al.).  There we showed that orders that remove more than the depth at the opposite best quote are rare.  Thus from the point of view of price formation we can neglect large orders that penetrate more than one price level in the limit order book, and simply assume that each transaction removes a limit order from the opposite best.  Although the size of orders ranges through more than four orders of magnitude, this variation is not an important effect in determining prices.
\item
{\it Stability of the order book}.  We require that there always be at least two orders on each side of the order book.  This ensures a well-defined sequence of prices\footnote{For the real data we sometimes observe situations where this condition is violated.  Though this assumption is somewhat {\it ad hoc}, we find that as well as making the simulations easier to perform, it improves the quality of our results.}. 
\end{itemize}

The simulation for a given stock is based on the parameter values in Table~\ref{paramSummary}.  Each time step of the simulation corresponds to the generation of a new trading order.   The order sign\footnote{Note that we are generating order signs exogenously.  As described in Section~\ref{orderSigns}, this is consistent with the assumption of Lillo, Mike, and Farmer (2005) that hidden order arrival is exogenous to price formation.}  is generated using a fractional gaussian process\footnote{In contrast to the more realistic model of order flow described in Section~\ref{orderSigns}, the fractional gaussian process does not allow us to control the prefactor of the correlation function, but rather generates a constant prefactor $C \approx 0.15$.  We find that this does not make much difference.} with Hurst exponent $H_s$, as described in Section~\ref{orderSigns}.  We generate an order price by drawing $x$ from a Student distribution with scale $\sigma_x$ and $\alpha_x$ degrees of freedom as described in Section~\ref{orderPlacement}.  If $x < s$ we generate a continuous approximation to the logarithmic price $\pi = x + \pi_b$ if it is a buy order or $\pi = \pi_a - x$ if it is a sell order.  This is then rounded to correspond to an integer tick price, i.e. the corresponding logarithmic price is specified by the relation $\exp(\pi_T) =  \mbox{int}(p/T)$, where $\mbox{int}(x)$ is the largest integer smaller than $x$.  Otherwise we place a market order and remove a limit order from the opposite best price; if this is the last order removed it causes a change in the midprice and the spread.  We decide which orders to cancel by generating random numbers according to the probability given by Equation~\ref{cancModel}.  The variable $y_i$ depends on the order $i$, so each order must be examined, and more than one order can be cancelled in a given time step.   The only exception is that as mentioned above we require that there always be at least two orders remaining on each side of the book, i.e. we do not cancel orders or allow transactions if this condition is not met.  
   
We initialize the limit order book with an arbitrary initial condition and run the simulation until it is approximately in a steady state\footnote{The initial state of the book is not important as long as we wait a sufficient length of time.  For the simulations described here we chose the initial book so that there are 10 orders on the best bid and 10 orders on the best ask, and ran the simulation for 10,000 iterations before sampling.}.   We then keep running the simulation to generate a series with twenty times more order placements than the real data sample.  The particular sequence of events generated in this manner depends on the random number seed used in the simulation, and will obviously not match the actual data in detail.  The comparison to the real data is therefore based only on the statistical properties of the prices.  For each sample we set the parameters to the appropriate value in Table~\ref{paramSummary}, run the simulation, measure the statistical properties of the price series as described below, and compare them to those of the real data.

\subsection{Comparison of simulated vs. real prices\label{simulatedVsReal}}

We test our model against real prices for all 25 samples described in Section~\ref{data}.  A summary of our results is shown in Table~\ref{comparison}.  For the volatility and the spread we compare the mean, standard deviation, and tail exponent of the prediction to that of the real data.
\begin{table}[htdp]
\begin{center}
\begin{tabular}{|l|l|l|l|l|l|ll|||||}
\hline
stock ticker&  $E(|r|) \times 10^{-4}$  & $E(s) \times 10^{-4} $ &  $\sigma(|r|)\times 10^{-4}$ & $\sigma(s) \times 10^{-4}$ & $\alpha(|r|)$ & $\alpha(s)$ \\
\hline
AZN & 5.4 $\pm$ 1.2 & 13.9 $\pm$ 0.6 & 7.2 $\pm$ 2.1 & 12.1 $\pm$ 0.7 & 2.4 $\pm$ 0.2 & 3.3 $\pm$ 0.3\\
predicted & 5.2 $\pm$ 0.1 & 13.8 $\pm$ 0.2 & 7.2 $\pm$ 0.3 & 11.9 $\pm$ 0.1 & 2.2 $\pm$ 0.4 & 3.2 $\pm$ 0.3\\
\hline
\hline
SHEL025 & 6.0 $\pm$ 0.8 & 13.4 $\pm$ 1.1 & 8.2 $\pm$ 0.8 & 8.8 $\pm$ 1.2 & 2.6 $\pm$ 0.5 & 3.7 $\pm$ 0.7\\
predicted & 6.4 $\pm$ 0.1 & 14.3 $\pm$ 0.2 & 6.2 $\pm$ 0.2 & 7.9 $\pm$ 0.2 & 2.5 $\pm$ 0.4 & 3.7 $\pm$ 0.2\\
\hline
\hline
PRU & 6.1 $\pm$ 0.9 & 17.3 $\pm$ 1.0 & 7.5 $\pm$ 0.7 & 11.1 $\pm$ 0.5 & 2.4 $\pm$ 0.5 & 2.9 $\pm$ 0.4\\
predicted & 6.9 $\pm$ 0.2 & 18.4 $\pm$ 0.3 & 12.9 $\pm$ 0.2 & 14.2 $\pm$ 0.3 & 2.4 $\pm$ 0.1 & 3.0 $\pm$ 0.2\\
\hline
\hline
REED & 7.3 $\pm$ 0.7 & 19.3 $\pm$ 1.1 & 7.2 $\pm$ 0.7 & 16.1 $\pm$ 0.3 & 2.3 $\pm$ 0.7 & 2.9 $\pm$ 0.6\\
predicted & 6.5 $\pm$ 0.2 & 18.8 $\pm$ 0.1 & 6.7 $\pm$ 0.2 & 16.2 $\pm$ 0.1 & 2.4 $\pm$ 0.1 & 2.8 $\pm$ 0.2\\
\hline
\hline
LLOY & 7.6 $\pm$ 1.3 & 17.1 $\pm$ 0.9 & 8.9 $\pm$ 0.6 & 13.8 $\pm$ 0.5 & 2.4 $\pm$ 0.4 & 3.5 $\pm$ 0.2\\
predicted & 7.3 $\pm$ 0.2& 16.9 $\pm$ 0.2 & 9.6 $\pm$ 0.2 & 13.2 $\pm$ 0.3 & 2.1 $\pm$ 0.3 & 3.2 $\pm$ 0.3\\
\hline
\hline
SHEL050 & 7.7 $\pm$ 1.1 & 17.3 $\pm$ 0.8 & 6.2 $\pm$ 0.7 & 9.6 $\pm$ 0.7 & 2.7 $\pm$ 0.5 & 3.9 $\pm$ 0.6\\
predicted & 7.4 $\pm$ 0.2 & 17.6 $\pm$ 0.3 & 5.7 $\pm$ 0.3 & 12.9 $\pm$ 0.3 & 2.8 $\pm$ 0.4 & 4.1 $\pm$ 0.2\\
\hline
\hline
SBRY & 8.3 $\pm$ 1.3 & 22.1 $\pm$ 1.1 & 8.2 $\pm$ 0.6 & 18.9 $\pm$ 0.8 & 2.3 $\pm$ 0.5 & 2.8 $\pm$ 0.5\\
predicted & 7.3 $\pm$ 0.2 & 19.8 $\pm$ 0.2 & 6.9 $\pm$ 0.2 & 19.7 $\pm$ 0.2 & 2.4 $\pm$ 0.1 & 2.9 $\pm$ 0.2\\
\hline
\hline
GUS & 8.8 $\pm$ 0.7 & 21.2 $\pm$ 1.2 & 7.2 $\pm$ 0.9 & 18.9 $\pm$ 0.8 & 2.3 $\pm$ 0.5 & 2.8 $\pm$ 0.5\\
predicted & 7.3 $\pm$ 0.2 & 19.8 $\pm$ 0.2 & 6.9 $\pm$ 0.2 & 18.1 $\pm$ 0.2 & 2.2 $\pm$ 0.2 & 2.7 $\pm$ 0.2\\
\hline
\hline
BSY050 & 9.4 $\pm$ 1.3 & 18.2 $\pm$ 1.5 & 8.8 $\pm$ 1.1 & 15.7 $\pm$ 1.3 & 2.4 $\pm$ 0.5 & 3.5 $\pm$ 0.3\\
predicted & 8.1 $\pm$ 0.1 & 16.7 $\pm$ 0.2 & 7.8 $\pm$ 0.2 & 14.9 $\pm$ 0.2 & 2.2 $\pm$ 0.2 & 3.3 $\pm$ 0.2\\
\hline
\hline 
BLT & 9.5 $\pm$ 0.9 & 23.1 $\pm$ 0.9 & 11.7 $\pm$ 1.6 & 21.1 $\pm$ 1.2 & 2.2 $\pm$ 0.7 & 2.8 $\pm$ 0.3\\
predicted & 8.2 $\pm$ 0.1 & 21.8 $\pm$ 0.1 & 10.3 $\pm$ 0.4  & 20.4 $\pm$ 0.3 & 2.4 $\pm$ 0.2 & 2.7 $\pm$ 0.2\\
\hline
\hline
\hline
III100 & 10.1 $\pm$ 1.2 & 29.4 $\pm$ 1.4 & 9.1 $\pm$ 1.1 & 20.8 $\pm$ 1.2 & 2.3 $\pm$ 0.5 & 3.0 $\pm$ 0.4\\
predicted & 6.1 $\pm$ 0.2 & 14.2 $\pm$ 0.2 & 5.5 $\pm$ 0.2 & 10.1 $\pm$ 0.2 & 2.3 $\pm$ 0.1 & 2.9 $\pm$ 0.2\\
\hline
\hline
BOC050 & 13.2 $\pm$ 1.3 & 33.3 $\pm$ 1.2 & 18.6 $\pm$ 0.7 & 42.1 $\pm$ 1.2 & 2.1 $\pm$ 0.8 & 2.9 $\pm$ 0.6\\
predicted & 5.3 $\pm$ 0.2 & 13.4 $\pm$ 0.2 & 4.2 $\pm$ 0.2 & 10.2 $\pm$ 0.1 & 2.9 $\pm$ 0.1 & 3.5 $\pm$ 0.3\\
\hline
\hline
III050 & 13.4 $\pm$ 1.4 & 38.3 $\pm$ 1.3 & 10.3 $\pm$ 0.9 & 25.3 $\pm$ 1.5 & 2.3 $\pm$ 0.3 & 2.9 $\pm$ 0.4\\
predicted & 9.3 $\pm$ 0.1 & 20.3 $\pm$ 0.1 & 8.9 $\pm$ 0.2 & 18.5 $\pm$ 0.2 & 2.5 $\pm$ 0.1 & 3.1 $\pm$ 0.2\\
\hline
\hline
BSY100 & 13.5 $\pm$ 1.6 & 26.6 $\pm$ 1.1 & 9.9 $\pm$ 1.1 & 17.2 $\pm$ 1.0 & 2.4 $\pm$ 0.4 & 3.4 $\pm$ 0.3\\
predicted & 7.1 $\pm$ 0.1 & 15.2 $\pm$ 0.1 & 4.7 $\pm$ 0.2 & 8.6 $\pm$ 0.2 & 2.6 $\pm$ 0.1 & 3.8 $\pm$ 0.2\\
\hline
\hline
BOC100 & 14.4 $\pm$ 1.1 & 34.5 $\pm$ 0.9 & 15.6 $\pm$ 0.9 & 33.5 $\pm$ 1.3 & 2.3 $\pm$ 0.4 & 3.2 $\pm$ 0.4\\
predicted & 6.4 $\pm$ 0.2 & 16.2 $\pm$ 0.1 & 3.8 $\pm$ 0.3 & 9.3 $\pm$ 0.1 & 3.1 $\pm$ 0.1 & 3.6 $\pm$ 0.2\\
\hline
\hline
DEB & 19.6 $\pm$ 1.7 & 65.4 $\pm$ 1.3 &  27.7 $\pm$ 1.8 & 72.2 $\pm$ 1.3 & 1.7 $\pm$ 0.5 & 2.4 $\pm$ 0.5\\
predicted & 7.9 $\pm$ 0.3 & 18.2 $\pm$ 0.2 & 7.7 $\pm$ 0.2 & 18.8 $\pm$ 0.2 & 2.1 $\pm$ 0.2 & 2.7 $\pm$ 0.2\\
\hline
\hline
TATE & 21.3 $\pm$ 1.1 & 68.2 $\pm$ 1.1 &  25.5 $\pm$ 1.1 & 69.7 $\pm$ 1.4 & 1.8 $\pm$ 0.6 & 2.4 $\pm$ 0.5\\
predicted & 8.3 $\pm$ 0.2 & 19.4 $\pm$ 0.2 & 6.9 $\pm$ 0.2 & 16.8 $\pm$ 0.3 & 2.4 $\pm$ 0.2 & 2.9 $\pm$ 0.1\\
\hline
\hline
NEX & 22.1 $\pm$ 1.1 & 68.4 $\pm$ 1.4 &  32.5 $\pm$ 1.2 & 68.1 $\pm$ 0.9 & 1.9 $\pm$ 0.4 & 2.8 $\pm$ 0.6\\
predicted & 7.0 $\pm$ 0.1 & 17.8 $\pm$ 0.1 & 7.3 $\pm$ 0.2 & 17.7 $\pm$ 0.1 & 2.4 $\pm$ 0.2 & 3.0 $\pm$ 0.2\\
\hline
\hline
BPB & 22.8 $\pm$ 1.4 & 69.1 $\pm$ 1.9 &  31.4 $\pm$ 1.9 & 71.9 $\pm$ 1.7 & 1.8 $\pm$ 0.4 & 2.5 $\pm$ 0.4\\
predicted & 8.5 $\pm$ 0.2 & 19.4 $\pm$ 0.2 & 6.9 $\pm$ 0.2 & 16.9 $\pm$ 0.2 & 2.2 $\pm$ 0.2 & 2.8 $\pm$ 0.2\\
\hline
 \end{tabular}
\caption{A comparison of statistical properties of the predictions (second row of each box) for the volatility $|r|$ and the spread $s$ to the real data (first row of each box) for Groups I (top ten) and II (bottom nine). The statistics are the sample mean $E$, the sample standard deviation $\sigma$, and the tail exponent $\alpha$.  Error bars are one standard deviation, computed using the variance plot method.  (Details can be provided on request) \label{comparison}}
\end{center}
\end{table}
The distribution for the spread is estimated by recording the best bid and ask prices immediately before order placements\footnote{The time when the spread is recorded can make a difference in the distribution.  The spread tends to narrow after receipt of limit orders and tends to widen after market orders or cancellations.}.

We find that the results cluster sharply into three groups.  Group I consists of the ten samples that have low volatility and low tick size, Group II consists of the nine samples with high volatility and low tick size, and Group III consists of the six samples with high tick size.  Low volatility means having average absolute transaction-to-transaction returns (based on the midprice) of less than $10^{-3}$, i.e. a tenth of a percent.  The threshold for separating large and small tick size is related to the ratio of the average price to the tick size, but it is more precisely determined by the stability properties of the model, as discussed in Section~\ref{stability}.  
 
\begin{figure}[ptb]
\begin{center}
\includegraphics[scale=0.4]{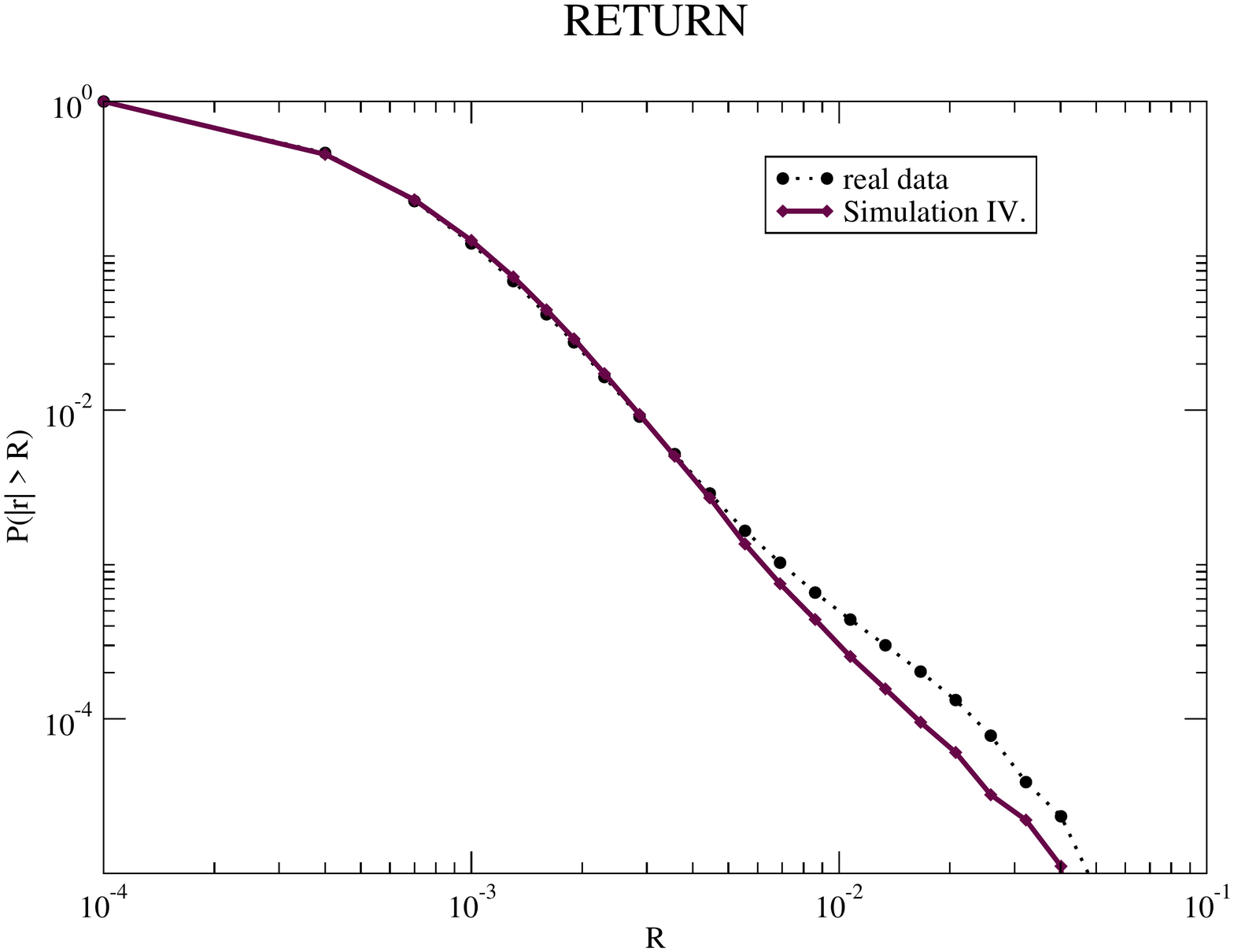}
\includegraphics[scale=0.4]{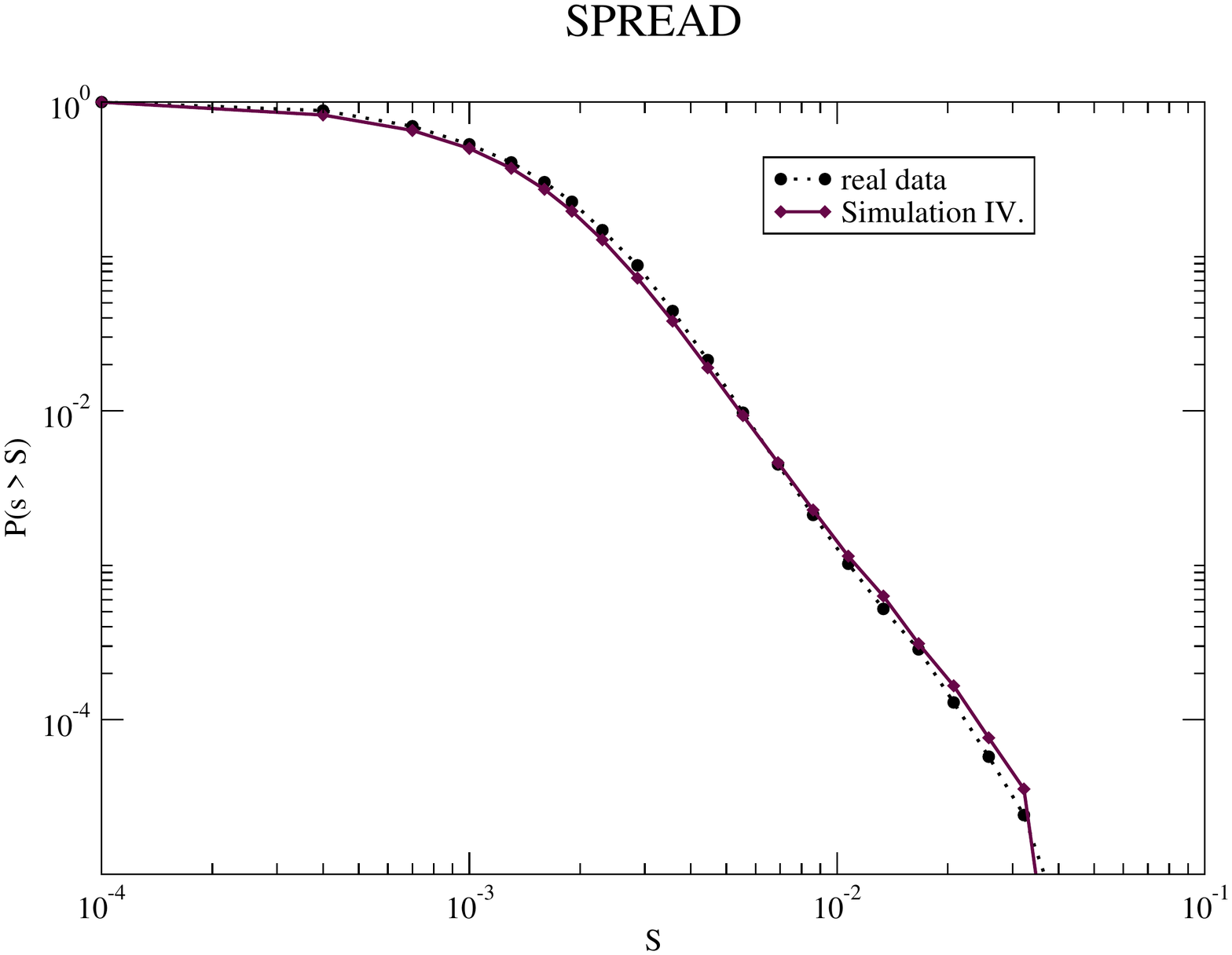}
\end{center}
\caption{A comparison of the distribution of predicted and actual volatility $|r|$ (upper) and spread $s$ (lower) for the stock Astrazeneca.  The solid curve is based on a single run of the model of length equal to the length of the data set, in this case 2,329,110 order placements.} 
\label{returnDist}
\end{figure}

For all the samples in Group I we find that the predictions are very good.  This is evident in Table~\ref{comparison}, where samples are ranked in order of volatility.  For Group I (the top ten rows), for most samples the predicted means of the return and the spread are within one standard deviation, for a couple they are within two standard deviations, and for one stock (GUS) they are slightly more than two standard deviations.  The statistical analysis becomes more complicated when one takes into account that the predictions are simulations and also have error bars; see the discussion a little later in this section.  We give a visual illustration of the correspondence between the predicted and actual distributions for spread and returns of a typical Group I stock in Figure~\ref{returnDist}.  The agreement is extremely good, both in terms of magnitude and functional form.

\begin{figure}[ptb]
\begin{center}
\includegraphics[scale=0.4]{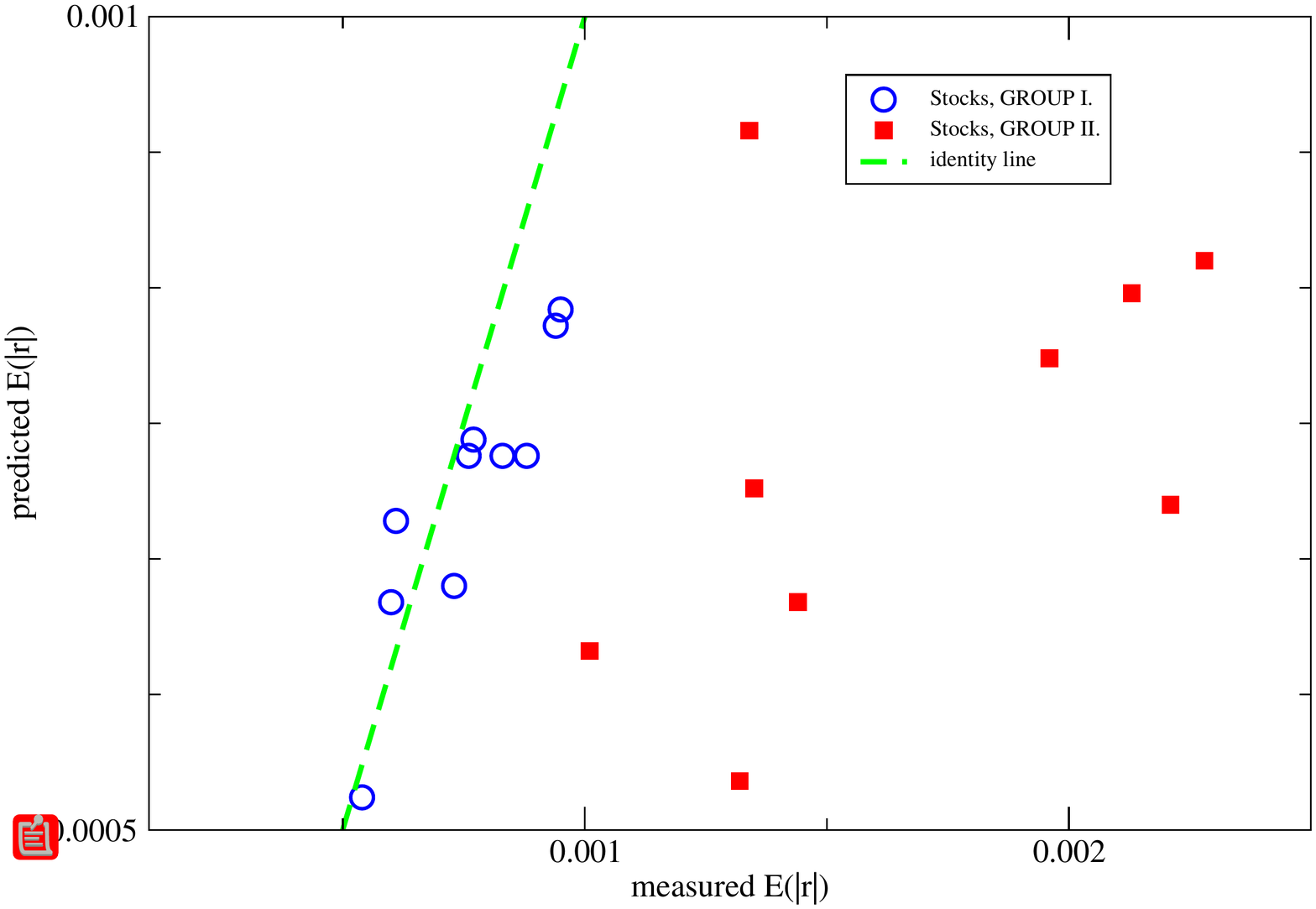}
\end{center}
\caption{The predicted volatility $E[|r|]$ is plotted against the actual volatility for samples in Group I (blue circles) and Group II (red squares).} 
\label{predComp}
\end{figure}

For Group II stocks, in contrast, the average predicted volatility and spread are consistently lower than the true values, in some cases by a large margin.  To make this more visually apparent, in Figure~\ref{predComp} we plot the predicted volatility against the actual volatility. 
We see that the predictions are quite good for Group I, but they get dramatically worse as soon as the volatility increases above $10^{-3}$, the threshold that defines the transition to Group II.  Even within Group I there is a tendency for the predictions to be somewhat low for the higher volatility stocks within the group, illustrating that while the model is good for Group I it is not perfect.

For the Group III stocks the simulation blows up, in the sense that the order book becomes infinitely full of orders and the predicted volatility goes to zero.   The stocks for which this happens are TSCO, VOD, HAS, NFDS, FGP, and AVE.  The reasons why this occurs are interesting for their own sake and are discussed in detail in the next section.

\subsection{Caveats}

We have only reported results for the distributions of returns and spreads.   There are many other properties that one could study, such as clustered volatility.   While the model displays some clustered volatility, it is weaker and less persistent than the real data.  For example, for AZN the Hurst exponent of volatility of the model is $H_v = 0.64$, in contrast to $H_v = 0.78$ for the real data.   Another area where the model fails is efficiency.  Autocorrelations in returns should be sufficiently close to zero that profits based on a linear extrapolation of returns are not possible.   For this model the autocorrelation function of returns drops to zero slower than the real data.   This is because, in the interest of keeping the model simple, there is no mechanism to adjust the liquidity for buying or selling in response to the imbalances of buying or selling that are driving the long-memory.     Finally, the fact that we had to introduce the {\it ad hoc} requirement that we preserve at least two orders in each side of the limit order book indicates that our existing order flow models have not fully captured the order book dynamics.

Despite these caveats, for Group I stocks the model does an extremely good job of describing the distribution of both returns and spreads.  We want to stress that these predictions are made without any adjustment of parameters based on formed prices.  All the parameters of the model are based on the order flow process alone -- there are no adjustable parameters to match the scale of the target data set.  Of course, causality can flow in both directions.  The parameters of the order flow process, particularly $\sigma_x$,  may be caused by properties of prices such as volatility.  In fact, Zovko and Farmer (\citeyear{Zovko02}) showed that in a study only of orders placed inside the book, the width of the distribution for order placement varies and tends to lag volatility.  The approximation we have made here averages over this effect (which may also contribute to reducing volatility fluctuations).

\subsection{Effect of tick size on model stability\label{stability}}

In this section we explain why the present model fails for large tick size stocks.  The problem comes from the unusual properties of the cancellation model constructed based on data from AZN, as discussed in Section~\ref{cancellationNumberOrders}.  There we showed that the probability of cancellation per order depends inversely on the total number of orders $n_{tot}$, and made the approximation that it is proportional to $1/n_{tot}$.  This is equivalent to saying that the total probability of cancellation (summed over all orders) is independent of the number of orders in the book.  This is a highly unexpected result.  In contrast, in the zero intelligence model (Daniels et al. \citeyear{Daniels03}) order cancellation was treated as  a Poisson process, so that the probability of cancellation of a given order is constant and the total probability of cancellation is proportional to $n_{tot}$.

This raises the question of how $n_{tot}$ can ever approach a reasonable steady state in the first place.  For the time average of the number of orders in the book $\langle n_{tot} \rangle$ to remain in the range $0 < \langle n_{tot} \rangle < \infty$, on average the order removal rate due to cancellations and transactions has to match the order deposit rate by limit order placement.  If the total cancellation is independent of the number of orders it has no influence on $\langle n_{tot} \rangle$.  The sole stabilizing force comes from the dependence of the transaction rate on the spread and the dependence of the spread on $n_{tot}$.  All else being equal, when $n_{tot}$ is small we expect the spread to be large, and vice versa.  To see how the stability mechanism works suppose the spread is large.  As demonstrated in Figure~\ref{transactionRatio}, this implies that the probability of a transaction is small, i.e. the deposition of a new limit order is much more likely than the removal of an order due to a transaction. (Remember that by definition each time unit corresponds to an order placement).  When $n_{tot}$ becomes small the spread becomes large, the transaction rate drops, and $n_{tot}$ increases.  Conversely, if $n_{tot}$ becomes large, the spread becomes small, the transaction rate becomes large, and $n_{tot}$ decreases.  From Figure~\ref{transactionRatio}, in the limit as the spread goes to zero the transaction rate approaches $1/2$, in contrast to the average transaction rate which is roughly $0.15$.  

\begin{figure}[ptb]
\begin{center}
\includegraphics[scale=0.6]{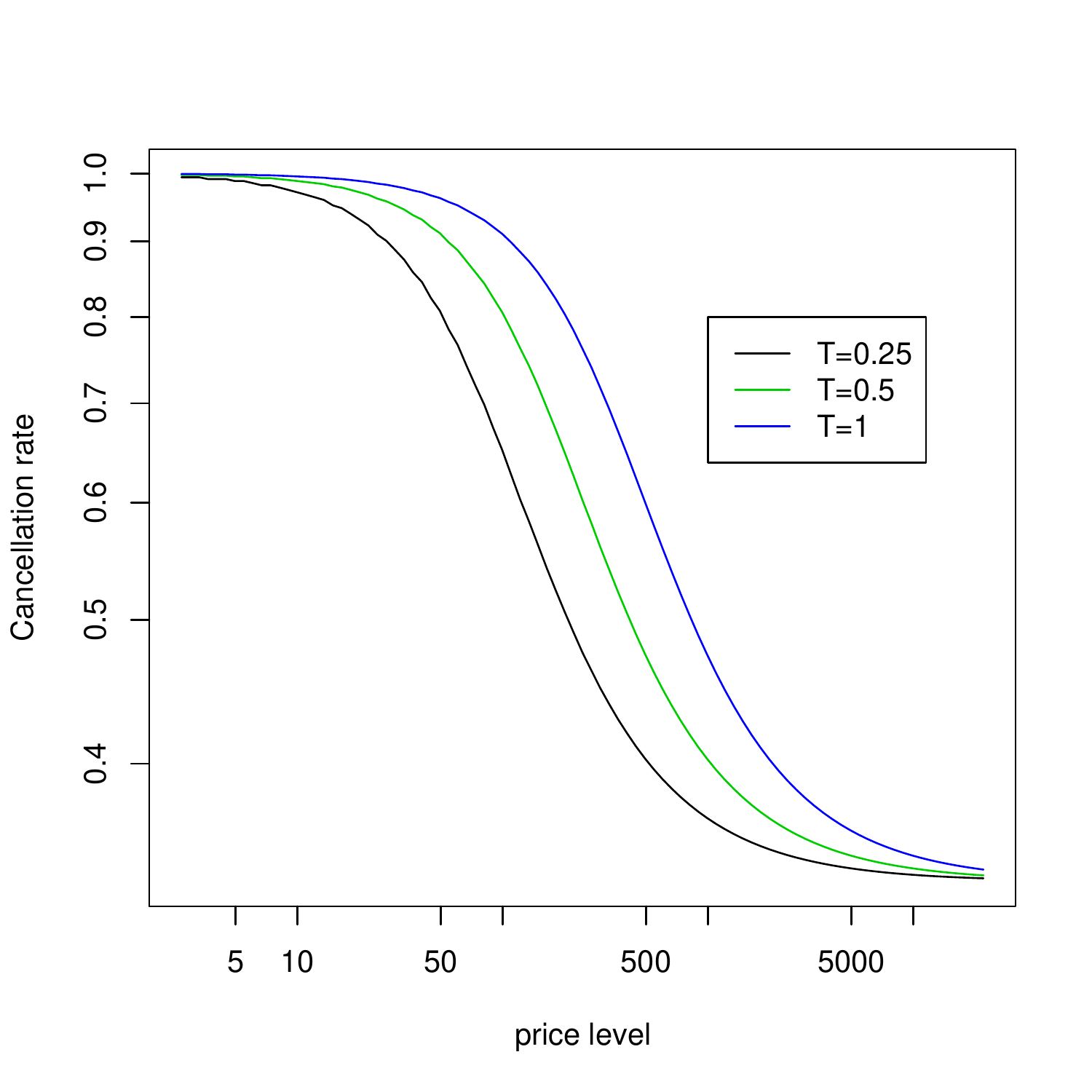}
\caption{Stability of the total number of orders in the book $n_{tot}$ with finite tick size.  In simulations of the model we vary the average price level $\langle p \rangle$ (horizontal axis) and the parameter $A$ (vertical axis), which is proportional to the cancellation rate.  We hold all the other parameters constant but use three different tick sizes $T$, corresponding to the three curves.  For parameter values in the upper right part of the diagram the average number of orders in the book remains bounded in $0 < \langle n_{tot} \rangle < \infty$, but to the lower left $\langle n_{tot} \rangle \to \infty$, which causes the average volatility to to go zero.}
\label{stabilityPlot}
\end{center}
\end{figure}

Thus, for small tick size stocks (Group I or II) the dynamical interaction between the spread, the total number of orders, and the transaction ratio keep the order book stable.  For large tick size stocks, however, the stabilizing mechanism is blocked by the fact that the spread can never be smaller than one tick.  If this lowers the upper limit on the transaction rate too much, orders accumulate in the book and $n_{tot}$ grows without bound.  This is illustrated in Figure~\ref{stabilityPlot}.  We use the parameter $A$ as a proxy for the overall cancellation rate.  We sweep $A$ and the average price $\langle p \rangle$ for three different tick sizes $T$, setting all the other parameters to be those of AZN\footnote{This illustrates that there is another implicit parameter in our model, which is the initial price.  The price process in our model is nonstationary, but over the three year time scale we simulate here, the price only changes by less than a factor of two.  In the real market this is roughly the time scale for stock splits and changes in the tick size.}.  For any given tick size, parameter values on the upper right side of the stability threshold curves produce order books with a well defined value of the time average $0 < \langle n_{tot} \rangle < \infty$, whereas for those on the lower left $\langle n_{tot} \rangle \to \infty$.  All the stocks in Groups I and II are in the upper right, whereas Group III (by definition) corresponds to those in the lower left part of the diagram.

\subsection{What causes the heavy tails of returns?\label{heavyTails1}}

\begin{figure}[tb]
\begin{center}
\includegraphics[scale=0.5]{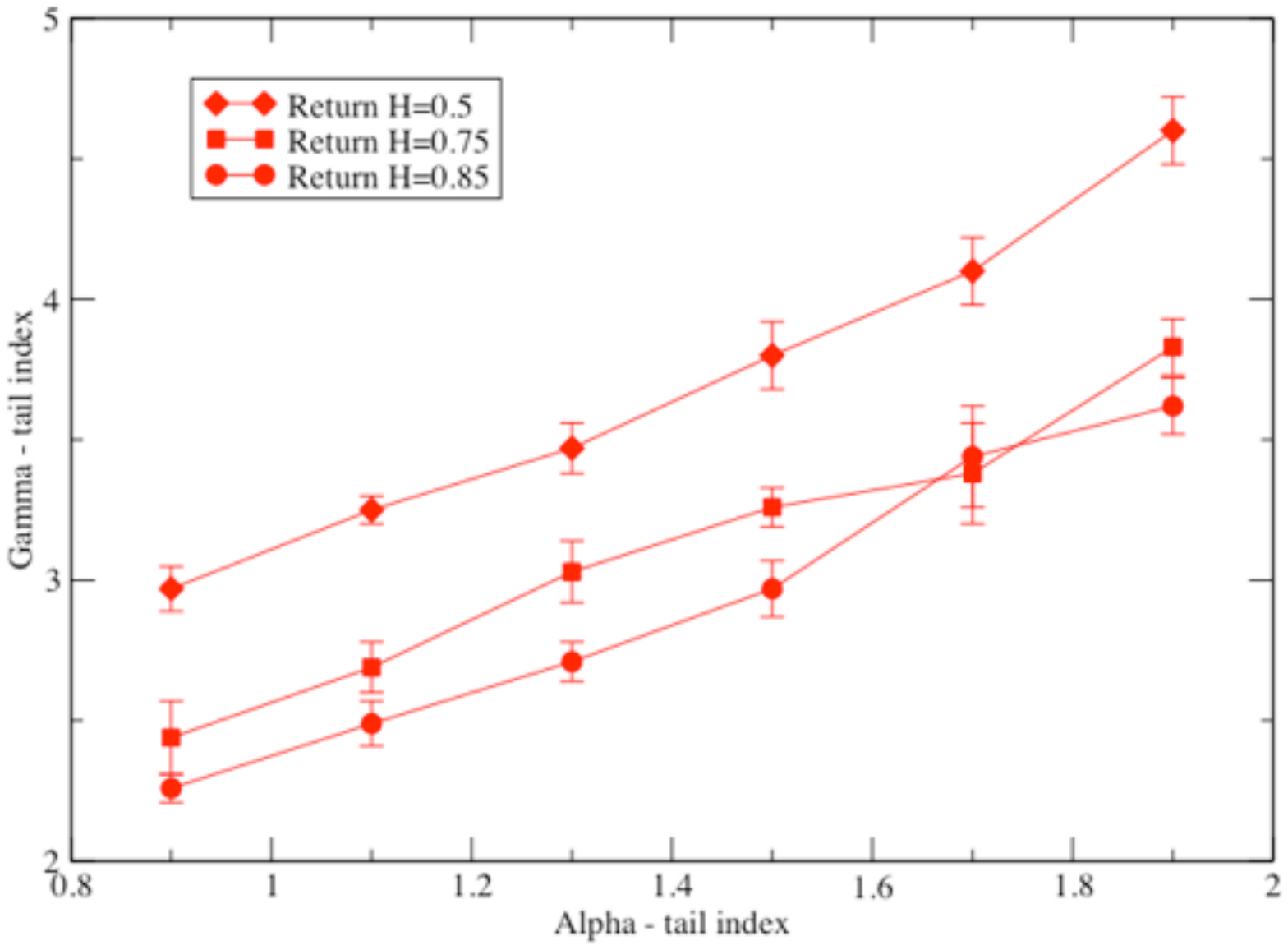}
\end{center}
\caption{Dependence of the tail exponent $\alpha_r$ of the volatility distribution on model parameters.  On the horizontal axis we vary the tail exponent $\alpha_x$ of the order price model, and on the vertical axis we plot estimates of the tail exponent $\alpha_r$ of the volatility distribution based on a Hill estimator.  We do this for three different values of the Hurst exponent $H_s$ of the order sign generation process.  All other parameters are those of AZN.  The tail exponent of the volatility distribution clearly depends on the parameters of the model.} 
\label{tailExponents}
\end{figure}

\begin{figure}[tb]
\begin{center}
\includegraphics[scale=0.4]{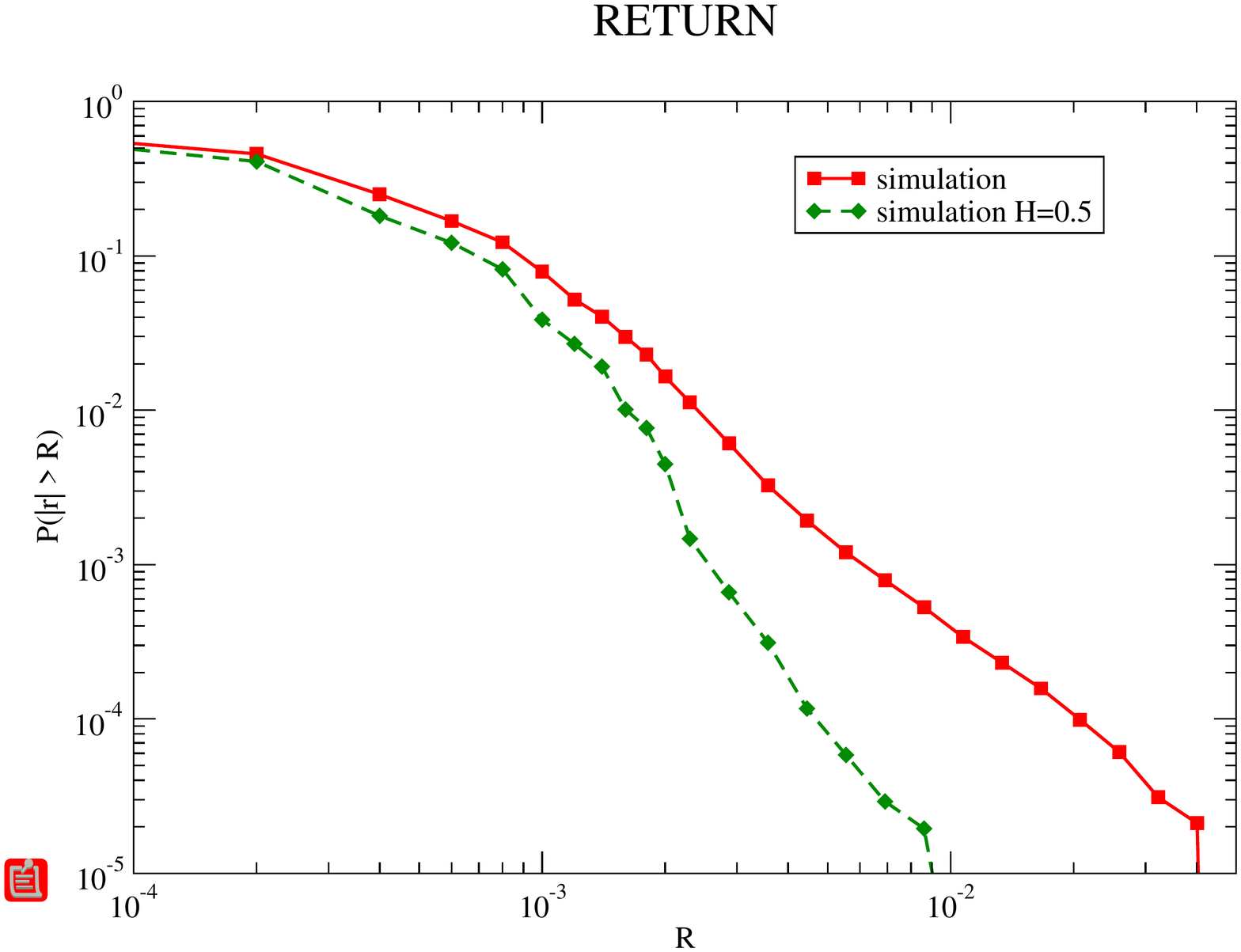}
\end{center}
\caption{A comparison of the distribution of volatility $P(|r| > R)$ for the model using a long-memory sign generation process (red squares) and an IID sign generation process (green diamonds).  All other parameters are held fixed to be those of AZN.} 
\label{returnComparison}
\end{figure}

Our model suggests that, at least at short time scales and for Group I stocks, the heavy tails of price returns are driven by market microstructure effects.  They depend both on the order sign and order placement process.  To study the dependence on microstructure effects more systematically in Figure~\ref{tailExponents} we vary the parameter $\alpha_x$, the tail exponent of the order placement distribution $P(x)$, for three different values of the Hurst exponent $H_s$ of the sign generation process: $H_s = 0.5$, $0.75$, and $0.85$.  We generate a series of a million order placements and measure the tail exponent $\alpha_r$ of the volatility $| r |$ using a Hill estimator.  The results make it clear that $\alpha_r$ depends on both parameters.  For $H_s = 0.85$, for example, as $\alpha_x$ is swept from $\alpha_x = 0.9$ to $1.9$, $\alpha_r$ varies from roughly $\alpha_r  = 2.2$ to $\alpha_r = 3.5$.  When we turn off the long-memory of the sign process by using $H_s = 0.5$ the heavy tails become much weaker; over the same range of variation of $\alpha_x$, $\alpha_r$ varies from roughly $\alpha_r = 3$ to $\alpha_r = 4.5$.
Figure~\ref{returnComparison} shows a comparison of the simulated return distribution with and without a long-memory process to generate the signs..    
The resulting tails of the return distribution are clearly much thinner, and no longer match those of the real data.


While our model is not a first principles explanation of why large returns follow a power law, it nonetheless strongly supports this hypothesis.  Since the model assumes two power law inputs, one for the temporal behavior of order signs, and one for the distribution of order placement prices, it is not surprising that it produces a power law, at least for large returns.  By assuming the inputs are power laws, we can perform arbitrarily long simulations of the model and thereby demonstrate that the large returns follow a power law, at a high level of statistical significance.  Thus, the question of whether returns are power laws depends on the question of how strong the evidence is that the inputs are power laws.  This evidence is strong.  As demonstrated by Lillo and Farmer (\citeyear{Lillo03c}) the evidence for long-memory in order signs is overwhelming, and as demonstrated by Zovko and Farmer (\citeyear{Zovko02}), who studied order placement inside the book for many stocks, the evidence for asymptotic convergence to a power law for $P^*(x)$ for large $|x|$ is also very good.   Thus, by showing a consistency of behavior, and by making it possible to test for power laws on alternative data, our models adds strength to the hypothesis that returns are power law distributed for large $|r|$, at least for the stocks we have studied.

We have not yet studied what happens when we aggregate returns at longer time scales.  It is well known that the property of having a power law tail with exponent $\alpha$ is preserved under IID aggregation.  When two different power laws are combined, the power law with the lowest exponent dominates.  It would require a highly unusual aggregation process to destroy this at longer time scales.  Thus this model suggests that the power laws seen in this model should persist at longer time scales, with tail exponents at least as small as those seen here.

       
\section{Concluding discussion}

We have built an empirical behavioral model for order placement that allows us to study the endogenous dynamics of liquidity and price formation in the order book.  It can be viewed as an agent based model, but it differs from most agent-based models in that the specification of the agents is quite simple and each component of the model is quantitatively grounded on empirical observations.   For the low volatility, small tick size stocks in our sample (which we call Group I), measurements of a small set of parameters of order flow give accurate predictions of the magnitude and functional form of the distribution of volatility and the spread.

Our model suggests that there is an equation of state linking the properties of order flow to the properties of prices.  By this we mean that there are constraints between the statistical properties of order flow and the statistical properties of prices, so that knowing one set of parameters automatically implies the other.    To see why we say this, please refer to Figure~\ref{predComp}, where we plot the volatility predicted by the model against the actual volatility.  The prediction of volatility varies because the order flow parameters in Table~\ref{comparison} vary.  The fact that there is agreement between the predicted and actual values for Group I shows that for these stocks the order flow parameters are sufficient to describe the volatility.  Of course, at this stage the equation of state remains implicit -- while the model captures it, we do not know how to explicitly write it down.  

This model shows how market microstructure effects, such as long-memory in the signs of orders, and heavy tails in the relative prices of orders in an auction, can generate heavy tails in price returns that closely match the data.  As discussed in the previous section, this reinforces the hypothesis that large returns asymptotically scale as a power law.  It also means that in order for the tail exponents of price returns to have a universal value near three, as previously hypothesized (Liu et al. \citeyear{Liu99}, Gabaix et al. \citeyear{Gabaix03,Gabaix06}), there must be constraints on the microstructure properties that enforce this.

The methodological approach that we have taken here can be viewed as a divide and conquer strategy.  We have tackled the problem of price formation by starting in the middle.  Rather than trying to immediately derive a model based on strategic motivations, we have empirically characterized behavioral regularities in order flow.  From here one can work in two directions, either working forward to understand the relation between order flow and price formation, or working backward to understand the strategic motivations that give rise to the regularities in the first place.  Here we have addressed the much easier problem of going forward, but our results are also potentially very useful for going backward.  It is always easier to solve a problem when it can be decomposed into pieces.  Going all the way from strategic motivations to prices is a much bigger step than moving from strategic motivations to regularities in order flow.  By empirically observing regularities in order flow we have created intermediate mileposts that any theory of strategic motivations should explain; once these are explained, we have shown that many features of prices follow more or less immediately.  At this point it is not obvious whether these  regularities can be explained in terms of rational choice, or whether they represent an example of irrational behavior, that can only be explained in terms of human psychology.

\bibliographystyle{authordate4}
\bibliography{jdf} 

\begin{thebibliography}{}

\bibitem[\protect\citename{Akgiray {\em et~al.}, }1989]{Akgiray89}
{\sc Akgiray, V., Booth, G.~G., \& Loistl, O.} 1989.
\newblock Stable laws are inappropriate for describing german stock returns.
\newblock {\em Allegemeines statistisches}, {\bf 73}(2), 115--121.

\bibitem[\protect\citename{Arthur {\em et~al.}, }1997]{Arthur97}
{\sc Arthur, W.~B., Holland, J.~H., LeBaron, B., Palmer, R., \& Tayler, P.}
  1997.
\newblock Asset pricing under endogenous expectations in an artificial stock
  market.
\newblock {\em Pages  15--44 of:} {\sc Arthur, W.~B., Durlauf, S.~N., \& Lane,
  D.~H.} (eds), {\em The economy as an evolving complex system ii}.
\newblock Redwood City: Addison-Wesley.

\bibitem[\protect\citename{Bak {\em et~al.}, }1997]{Bak97}
{\sc Bak, P., Paczuski, M., \& Shubik, M.} 1997.
\newblock Price variations in a stock market with many agents.
\newblock {\em Physica a-statistical mechanics and its applications}, {\bf
  246}(3-4), 430--453.

\bibitem[\protect\citename{Barberis \& Thaler, }2003]{Barberis03}
{\sc Barberis, N., \& Thaler, R.} 2003.
\newblock A survey of behavioral finance.
\newblock {\em In:} {\sc Constantinides, G., Harris, M., \& Stultz, R.} (eds),
  {\em Handbook of economics and finance}.
\newblock North Holland.

\bibitem[\protect\citename{Beran, }1994]{Beran94}
{\sc Beran, J.} 1994.
\newblock {\em Statistics for long-memory processes}.
\newblock New York: Chapman \& Hall.

\bibitem[\protect\citename{Bollerslev {\em et~al.}, }1997]{Bollerslev97}
{\sc Bollerslev, T., Domowitz, I., \& Wang, J.} 1997.
\newblock Order flow and the bid-ask spread: An empirical probability model of
  screen-based trading.
\newblock {\em Journal of economic dynamics and control}, {\bf 21}(8-9),
  1471--1491.

\bibitem[\protect\citename{Bouchaud {\em et~al.}, }2002]{Bouchaud02}
{\sc Bouchaud, J-P., Mezard, M., \& Potters, M.} 2002.
\newblock Statistical properties of the stock order books: empirical results
  and models.
\newblock {\em Quantitative finance}, {\bf 2}(4), 251--256.

\bibitem[\protect\citename{Bouchaud {\em et~al.}, }2004]{Bouchaud04}
{\sc Bouchaud, J-P., Gefen, Y., Potters, M., \& Wyart, M.} 2004.
\newblock Fluctuations and response in financial markets: The subtle nature of
  ``random" price changes.
\newblock {\em Quantitative finance}, {\bf 4}(2), 176--190.

\bibitem[\protect\citename{Bouchaud {\em et~al.}, }2006]{Bouchaud04b}
{\sc Bouchaud, J-P., Kockelkoren, J., \& Potters, M.} 2006.
\newblock Random walks, liquidity molasses and critical response in financial
  markets.
\newblock {\em Quantitative finance}, {\bf 6}(2), 115--123.

\bibitem[\protect\citename{Brock \& Hommes, }1999]{Brock99}
{\sc Brock, W.~A., \& Hommes, C.~H.} 1999.
\newblock Rational animal spirits.
\newblock {\em Pages  109--137 of:} {\sc Herings, P.J.J., van~der Laan, G., \&
  Talman, A.J.J.} (eds), {\em The theory of markets}.
\newblock Amsterdam: North Holland.

\bibitem[\protect\citename{Camerer {\em et~al.}, }2003]{Camerer03}
{\sc Camerer, C.F., Loewenstein, G., \& Rabin, M.} (eds). 2003.
\newblock {\em Advances in behavioral economics}.
\newblock Princeton University Press.

\bibitem[\protect\citename{Challet \& Stinchcombe, }2001]{Challet01}
{\sc Challet, D., \& Stinchcombe, R.} 2001.
\newblock Analyzing and modeling 1+1d markets.
\newblock {\em Physica a}, {\bf 300}(1-2), 285--299.

\bibitem[\protect\citename{Challet \& Stinchcombe, }2003]{Challet02}
{\sc Challet, D., \& Stinchcombe, R.} 2003.
\newblock Nonconstant rates and overdiffusive prices in simple models of limit
  order markets.
\newblock {\em Quantitative finance}, {\bf 3}, 165.

\bibitem[\protect\citename{Challet {\em et~al.}, }2005]{Challet05}
{\sc Challet, Damien, Marsili, Matteo, \& Zhang, Yi-Cheng}. 2005.
\newblock {\em Minority games}.
\newblock Oxford: Oxford University Press.

\bibitem[\protect\citename{Chang {\em et~al.}, }2002]{Chang02}
{\sc Chang, Iksoo, Stauffer, Dietrich, \& Pandey, Ras~B}. 2002.
\newblock Asymmetries, correlations and fat tails in percolation market model.
\newblock {\em International journal of theoretical and applied finance}, {\bf
  5}(6), 585--97.

\bibitem[\protect\citename{Chiarella \& Iori, }2002]{Chiarella02}
{\sc Chiarella, C., \& Iori, G.} 2002.
\newblock A simulation analysis of the microstructure of double auction
  markets.
\newblock {\em Quantitative finance}, {\bf 2}, 346--353.

\bibitem[\protect\citename{Cohen {\em et~al.}, }1985]{Cohen85}
{\sc Cohen, K.~J., Conroy, R.~M., \& Maier, S.~F.} 1985.
\newblock Order flow and the quality of the market.
\newblock {\em Pages  93--110 of:} {\sc Amihud, Y., Ho, T.S.Y, \& Schwartz,
  R.~A} (eds), {\em Market making and the changing structure of the securities
  industry}.
\newblock Lanham: Rowman \& Littlefield.

\bibitem[\protect\citename{Daniels {\em et~al.}, }2003]{Daniels03}
{\sc Daniels, M.~G., Farmer, J.~D., Gillemot, L., Iori, G., \& Smith, D.~E.}
  2003.
\newblock Quantitative model of price diffusion and market friction based on
  trading as a mechanistic random process.
\newblock {\em Physical review letters}, {\bf 90}(10), 108102--108104.

\bibitem[\protect\citename{Demsetz, }1968]{Demsetz68}
{\sc Demsetz, Harold}. 1968.
\newblock The cost of transacting.
\newblock {\em The quarterly journal of economics}, {\bf 82}, 33--53.

\bibitem[\protect\citename{Domowitz \& Wang, }1994]{Domowitz94}
{\sc Domowitz, I., \& Wang, J.} 1994.
\newblock Auctions as algorithms: computerized trade execution and price
  discovery.
\newblock {\em Journal of economic dynamics and control}, {\bf 18}(1), 29--60.

\bibitem[\protect\citename{Easley \& O'Hara, }1992]{Easley92}
{\sc Easley, D., \& O'Hara, M.} 1992.
\newblock Time and the process of security price adjustment.
\newblock {\em The journal of finance}, {\bf 47}(2), 577--605.

\bibitem[\protect\citename{Eliezer \& Kogan, }1998]{Eliezer98}
{\sc Eliezer, D., \& Kogan, I.~I.} 1998.
\newblock {\em Scaling laws for the market microstructure of the interdealer
  broker markets}.
\newblock Tech. rept. http://www.arxiv.org/abs/cond-mat/9808240.

\bibitem[\protect\citename{Fama, }1965]{Fama65}
{\sc Fama, E.~F.} 1965.
\newblock The behavior of stock-market prices.
\newblock {\em The journal of business}, {\bf 38}(1), 34--105.

\bibitem[\protect\citename{Farmer \& Lillo, }2004]{Farmer04}
{\sc Farmer, J.~D., \& Lillo, F.} 2004.
\newblock On the origin of power laws in financial markets.
\newblock {\em Quantitative finance}, {\bf 314}, 7--10.

\bibitem[\protect\citename{Farmer {\em et~al.}, }2004]{Farmer04b}
{\sc Farmer, J.~D., Gillemot, L., Lillo, F., Mike, S., \& Sen, A.} 2004.
\newblock What really causes large price changes?
\newblock {\em Quantitative finance}, {\bf 4}(4), 383--397.

\bibitem[\protect\citename{Farmer {\em et~al.}, }2005]{Farmer05}
{\sc Farmer, J.~D., Patelli, P., \& Zovko, Ilija}. 2005.
\newblock The predictive power of zero intelligence in financial markets.
\newblock {\em Proceedings of the national academy of sciences of the united
  states of america}, {\bf 102}(6), 2254--2259.

\bibitem[\protect\citename{Farmer {\em et~al.}, }2006]{Farmer06}
{\sc Farmer, J.D., Gerig, A., Lillo, F., \& Mike, S.} 2006.
\newblock Market efficiency and the long-memory of supply and demand: Is price
  impact variable and permanent or fixed and temporary?
\newblock {\em Quantitative finance}, {\bf 6}(2), 107--112.

\bibitem[\protect\citename{Foucault {\em et~al.}, }2005]{Foucault01}
{\sc Foucault, T., Kadan, O., \& Kandel, E.} 2005.
\newblock Limit order book as a market for liquidity.
\newblock {\em The review of financial studies}, {\bf 18}(4), 1171--1217.

\bibitem[\protect\citename{Gabaix {\em et~al.}, }2003]{Gabaix03}
{\sc Gabaix, X., Gopikrishnan, P., Plerou, V., \& Stanley, H.~E.} 2003.
\newblock A theory of power-law distributions in financial market fluctuations.
\newblock {\em Nature}, {\bf 423}, 267--270.

\bibitem[\protect\citename{Gabaix {\em et~al.}, }2006]{Gabaix06}
{\sc Gabaix, X., Gopikrishnan, P., Plerou, V., \& Stanley, H.E.} 2006.
\newblock Institutional investors and stock market volatility.
\newblock {\em Quarterly journal of economics}, {\bf 121}, 461--504.

\bibitem[\protect\citename{Gell-Mann \& Tsallis, }2004]{Gell-Mann04}
{\sc Gell-Mann, Murray, \& Tsallis, Constantino} (eds). 2004.
\newblock {\em Nonextensive entropy-interdisciplinary applications}.
\newblock SFI Studies in the Sciences of Complexity.
\newblock New York: Oxford University Press.

\bibitem[\protect\citename{Giardina \& Bouchaud, }2003]{Giardina03}
{\sc Giardina, I., \& Bouchaud, J-P.} 2003.
\newblock Bubbles, crashes and intermittency in agent based market models.
\newblock {\em European physical journal b}, {\bf 31}(3), 421--437.

\bibitem[\protect\citename{Gillemot {\em et~al.}, }2006]{Gillemot05}
{\sc Gillemot, Laszlo, Farmer, J.~Doyne, \& Lillo, Fabrizio}. 2006.
\newblock There's more to volatility than volume.
\newblock {\em Quantitative finance}, {\bf 6}(5), 371--384.

\bibitem[\protect\citename{Glosten, }1988]{Glosten88}
{\sc Glosten, L.~L.} 1988.
\newblock Estimating the components of the bid/ask spread.
\newblock {\em Journal of financial economics}, {\bf 21}, 123--142.

\bibitem[\protect\citename{Glosten, }1992]{Glosten92}
{\sc Glosten, Lawrence}. 1992.
\newblock {\em Equilibrium in an electronic open limit order book}.
\newblock Copies available from: Columbia University, Graduate School of
  Business, First Boston Series, New York, NY 10027.

\bibitem[\protect\citename{Goldstein {\em et~al.}, }2004]{Goldstein04b}
{\sc Goldstein, M.~L., Morris, S.~A., \& Yen, G.~G.} 2004.
\newblock Problems with fitting to the power-law distribution.
\newblock {\em the european physical journal b}, {\bf 41}, 255--258.

\bibitem[\protect\citename{Gopikrishnan {\em et~al.}, }2000]{Gopikrishnan00}
{\sc Gopikrishnan, P., Plerou, V., Gabaix, X., \& Stanley, H.~E.} 2000.
\newblock Statistical properties of share volume traded in financial markets.
\newblock {\em Physical review e}, {\bf 62}(4), R4493--R4496.
\newblock Part A.

\bibitem[\protect\citename{Hirschleifer, }2001]{Hirschleifer01}
{\sc Hirschleifer, D.} 2001.
\newblock Investor psychology and asset pricing.
\newblock {\em Journal of finance}, {\bf 56}(4), 1533--1597.

\bibitem[\protect\citename{Koedijk {\em et~al.}, }1990]{Koedijk90}
{\sc Koedijk, K.~G., Schafgans, M. M.~A., \& de~Vries, C.~G.} 1990.
\newblock The tail index of exchange rates.
\newblock {\em Journal of international economics}, {\bf 29}(1-2), 1--197.

\bibitem[\protect\citename{LeBaron, }2001]{LeBaron01b}
{\sc LeBaron, B.} 2001.
\newblock Empirical regularities from interacting long and short memory
  investors in an agent-based financial market.
\newblock {\em Ieee transactions on evolutionary computation}, {\bf 5},
  442--455.

\bibitem[\protect\citename{Lillo \& Farmer, }2004]{Lillo03c}
{\sc Lillo, F., \& Farmer, J.~D.} 2004.
\newblock The long memory of the efficient market.
\newblock {\em Studies in nonlinear dynamics \& econometrics}, {\bf 8}(3).

\bibitem[\protect\citename{Lillo \& Farmer, }2005]{Lillo05}
{\sc Lillo, F., \& Farmer, J.~D.} 2005.
\newblock The key role of liquidity fluctuations in determining large price
  fluctuations.
\newblock {\em Fluctuations and noise letters}, {\bf 5}, L209--L216.

\bibitem[\protect\citename{Lillo {\em et~al.}, }2005]{Lillo05b}
{\sc Lillo, F., Mike, S., \& Farmer, J.~D.} 2005.
\newblock Theory for long memory in supply and demand.
\newblock {\em Physical review e}, {\bf 7106}(6), 066122.

\bibitem[\protect\citename{Liu {\em et~al.}, }1999]{Liu99}
{\sc Liu, F., Gopikrishnan, P., Cizeau, P, Meyer, M., Peng, C.-K., \& Stanley,
  H.E.} 1999.
\newblock The statistical properties of the volatility of price fluctuations.
\newblock {\em Physical review e.}, {\bf 60}, 1390--1400.

\bibitem[\protect\citename{Longin, }1996]{Longin96}
{\sc Longin, F.~M.} 1996.
\newblock The asymptotic distribution of extreme stock market returns.
\newblock {\em The journal of business}, {\bf 69}(3), 383--408.

\bibitem[\protect\citename{Loretan \& Phillips, }1994]{Loretan94}
{\sc Loretan, M., \& Phillips, P. C.~B.} 1994.
\newblock Testing the covariance stationarity of heavy-tailed time series: An
  overview of the theory with applications to several financial datasets.
\newblock {\em Journal of empirical finance}, {\bf 1}(2), 211--248.

\bibitem[\protect\citename{Lux, }1996]{Lux96}
{\sc Lux, T.} 1996.
\newblock The stable paretian hypothesis and the frequency of large returns: an
  examination of major german stocks.
\newblock {\em Applied financial economics}, {\bf 6}(6), 463--475.

\bibitem[\protect\citename{Lux \& Marchesi, }1999]{Lux99}
{\sc Lux, T., \& Marchesi, M.} 1999.
\newblock Scaling and criticality in a stochastic multi-agent model of a
  financial market.
\newblock {\em Nature}, {\bf 397}(6719), 498--500.

\bibitem[\protect\citename{Mandelbrot, }1963]{Mandelbrot63}
{\sc Mandelbrot, B.} 1963.
\newblock The variation of certain speculative prices.
\newblock {\em The journal of business}, {\bf 36}(4), 394--419.

\bibitem[\protect\citename{Mantegna \& Stanley, }1995]{Mantegna95}
{\sc Mantegna, R.~N., \& Stanley, H.~E.} 1995.
\newblock Scaling behavior in the dynamics of an economic index.
\newblock {\em Nature}, {\bf 376}(6535), 46--49.

\bibitem[\protect\citename{Maslov, }2000]{Maslov00}
{\sc Maslov, S.} 2000.
\newblock Simple model of a limit order-driven market.
\newblock {\em Physica a-statistical mechanics and its applications}, {\bf
  278}(3-4), 571--578.

\bibitem[\protect\citename{Mendelson, }1982]{Mendelson82}
{\sc Mendelson, H.} 1982.
\newblock Market behavior in a clearing house.
\newblock {\em Econometrica}, {\bf 50}(6), 1505--1524.

\bibitem[\protect\citename{Muller {\em et~al.}, }1998]{Muller98}
{\sc Muller, U.~A., Dacorogna, M.~M., \& Pictet, O.~V.} 1998.
\newblock Heavy tails in high-frequency financial data.
\newblock {\em Pages  55--78 of:} {\sc Adler, R.~J., Feldman, R.~E., \& Taqqu,
  M.~S.} (eds), {\em A practical guide to heavy tails: Statistical techniques
  and applications}.
\newblock Berlin: Springer-Verlag.

\bibitem[\protect\citename{Officer, }1972]{Officer72}
{\sc Officer, R.~R.} 1972.
\newblock Distribution of stock returns.
\newblock {\em Journal of the american statistical association}, {\bf 67}(340),
  807--812.

\bibitem[\protect\citename{Peng {\em et~al.}, }1994]{Peng94}
{\sc Peng, C-K., Buldyrev, S.~V., Havlin, S., Simons, M., Stanley, H.~E., \&
  Goldberger, A.~L.} 1994.
\newblock Mosaic organization of dna nucleotides.
\newblock {\em Physical review e}, {\bf 49}(2), 1685--1689.

\bibitem[\protect\citename{Plerou {\em et~al.}, }1999]{Plerou99}
{\sc Plerou, V., Gopikrishnan, P., Amaral, L. A.~N., Meyer, M., \& Stanley,
  H.~E.} 1999.
\newblock Scaling of the distribution of price fluctuations of individual
  companies.
\newblock {\em Physical review e}, {\bf 60}(6), 6519--6529.
\newblock Part A.

\bibitem[\protect\citename{Potters \& Bouchaud, }2003]{Potters03}
{\sc Potters, M., \& Bouchaud, J-P.} 2003.
\newblock More statistical properties of order books and price impact.
\newblock {\em Physica a}, {\bf 324}, 133--140.

\bibitem[\protect\citename{Rachev \& Mittnik, }2000]{Rachev00}
{\sc Rachev, S., \& Mittnik, S.} 2000.
\newblock {\em Stable paretian models in finance}.
\newblock New York: John Wiley \& Sons.

\bibitem[\protect\citename{Sandas, }2001]{Sandas01}
{\sc Sandas, P.} 2001.
\newblock Adverse selection and comparative market making:empirical evidence
  from a limit order market.
\newblock {\em The review of financial studies}, {\bf 14}(3), 705--734.

\bibitem[\protect\citename{Schleifer, }2000]{Schleifer00}
{\sc Schleifer, A.} 2000.
\newblock {\em Clarendon lectures: Inefficient markets}.
\newblock Oxford: Oxford University Press.

\bibitem[\protect\citename{Slanina, }2001]{Slanina01}
{\sc Slanina, F.} 2001.
\newblock Mean-field approximation for a limit order driven market model.
\newblock {\em Physical review e}, {\bf 64}(5), article no.056136.
\newblock Part 2.

\bibitem[\protect\citename{Smith {\em et~al.}, }2003]{Smith03}
{\sc Smith, E., Farmer, J.~D., Gillemot, L., \& Krishnamurthy, S.} 2003.
\newblock Statistical theory of the continuous double auction.
\newblock {\em Quantitative finance}, {\bf 3}(6), 481--514.

\bibitem[\protect\citename{Stoll, }1978]{Stoll78}
{\sc Stoll, Hans~R}. 1978.
\newblock The supply of dealer services in securities markets.
\newblock {\em Journal of finance}, {\bf 33}(4), 1133--51.

\bibitem[\protect\citename{Tang \& Tian, }1999]{Tang99}
{\sc Tang, L.~H., \& Tian, G.~S.} 1999.
\newblock Reaction-diffusion-branching models of stock price fluctuations.
\newblock {\em Physica a}, {\bf 264}(3-4), 543--550.

\bibitem[\protect\citename{Thaler, }2005]{Thaler05}
{\sc Thaler, R.} 2005.
\newblock {\em Advances in behavioral economics}.
\newblock Russel Sage Foundation.

\bibitem[\protect\citename{Tsallis, }1988]{Tsallis88}
{\sc Tsallis, C.} 1988.
\newblock Possible generalization of boltzmann-gibbs statistics.
\newblock {\em Journal of statistical physics}, {\bf 52}(1-2), 479--487.

\bibitem[\protect\citename{Vaglica {\em et~al.}, }2007]{Vaglica06}
{\sc Vaglica, Lillo, F., Moro, E., \& Mantegna, R.} 2007.
\newblock {\em Scaling laws of strategic behavior and size heterogeneity in
  agent dynamics}.
\newblock Tech. rept. http://arxiv.org/pdf/0704.2003.

\bibitem[\protect\citename{Weber \& Rosenow, }2006]{Weber04}
{\sc Weber, P., \& Rosenow, B.} 2006.
\newblock Large stock price changes: volume or liquidity?
\newblock {\em Quantitative finance}, {\bf 6}(1), 7--14.

\bibitem[\protect\citename{White, }2006]{White06}
{\sc White, D.} 2006.
\newblock {\em Tutorial for using q-exponentials for city size distributions}.
\newblock Tech. rept. U.C. Irvine.

\bibitem[\protect\citename{Zovko \& Farmer, }2002]{Zovko02}
{\sc Zovko, I., \& Farmer, J.~D.} 2002.
\newblock The power of patience; a behavioral regularity in limit order
  placement.
\newblock {\em Quantitative finance}, {\bf 2}(5), 387--392.

\end{thebibliography}

\end{document}